\RequirePackage{lineno}
\documentclass[prd,aps,showpacs,preprintnumbers,amsmath,amssymb,floatfix,nofootinbib]{revtex4-2}
\usepackage{amssymb}
\usepackage{amsmath}    
\usepackage{graphicx}
\usepackage{xcolor}
\usepackage{wasysym}
\begin{document}
\title{A model for the curvature response of the CDF II drift chamber}
\author{Ashutosh Vijay Kotwal}
\affiliation{Department of Physics, Duke University, Durham, North Carolina 27708, USA}
\begin{abstract}
The CDF II experiment at the Fermilab Tevatron used a drift chamber to measure the momenta of charged particles. 
 We present a model for the response of the drift chamber to the curvature of a charged particle's trajectory.
Constraints on the model parameters are obtained from cosmic-ray data and from information published by CDF  in the
 context of the $W$ boson mass measurement. Implications for the calibration of the drift chamber measurement of 
 momentum  are discussed. The robustness of the CDF calibration procedure is demonstrated.
 The model provides a framework for the analysis of precision magnetic trackers of high-momentum particles. 
\end{abstract}
\maketitle
\section{Introduction}
\label{sec:intro}
\hspace*{0.06in}
 The CDF II collider detector operated at the Fermilab Tevatron during 1999-2011. 
 The central drift chamber of the CDF II experiment, called the central outer tracker (COT) measured the 
 positions of charged particles as they traverse the active volume~\cite{cotNim}. The cylindrical chamber was coaxial with the colliding beams and
 and was immersed in a 1.4~T axial magnetic field that bent charged particles in the plane transverse to the mutual axis. The particle positions 
 were recorded by up to 96 wires sequentially in the outward radial direction.  Half of the sense wires 
 were axial and the remaining wires had a small stereo angle to collectively determine the particles' positions and directions in 
 three dimensions.    A helical fit is performed to the measured coordinates
 as a function of radius to infer the track parameters. 

 In the context of precision measurements like that of the $W$ boson mass $m_W$, the CDF drift chamber is the crucial 
 device because it provides the most important input -- the measurement of the momenta of the muons and electrons 
 that originate from the decay of $W$ and $Z$ bosons and the $J/\psi$ and $\Upsilon$ mesons. The silicon vertex detector is not used 
 in the CDF measurements of $m_W$ because the improvement in the momentum resolution is marginal and there is no benefit to the analysis~\cite{CDF2022,CDF2firstPRD}. 

The helical trajectory of the particle in the axial magnetic field projects to a two-dimensional circle in the traverse plane. 
 The circle's radius $R$ is proportional to $p_T$, the momentum component transverse to the beam axis, and 
 its curvature $c$ is proportional to $q R^{-1}$,
 where $q = \pm1$ is the particle's electric charge\footnote{Experimentally, the particle's charge is determined from the sign of 
 $c^{\rm measured}$.}. Therefore, in the natural units\footnote{We  use natural units by setting the speed
  of light to unity. } of $p_T^{-1}$ (typically GeV$^{-1}$),  
 the curvature of the fitted track, $c^{\rm measured} \equiv (q/p_T)^{\rm measured}$ is a 
 function of the true curvature $c \equiv q/p_T$. This is the most important response function since it is associated with the momentum
 calibration of the device. 

 We begin this discussion with the ansatz that an open-cell drift chamber in which the entire volume is active,
 instrumented and fiducial must have an analytic response function for high-$p_T$ particles. 
 In this context, analyticity refers to smoothness, i.e. continuity and differentiability. Later 
 we expand the discussion to the implications of and constraints on non-analytic behavior.

 A salient feature of this drift chamber is that many of the terms in general ansatz for the response function can be related 
 to the fundamentals of the chamber's construction and operation. This feature enables the calibration of the device largely
 from first principles. Pushing this approach as far as possible maximizes the understanding of ``how and why it works'' 
  in the sense of reductionism and critical rationalism~\cite{popper} and avoids the black-box nature of very high-dimensional
 fitting or machine learning~\cite{interpretability}.  

Guided by this philosophy, the goal of this paper is to analyse the drift chamber's calibration for the 
 transverse momentum of charged particles. While the measurement of the polar angle is also important for the reconstruction of invariant
 masses from the 3-momenta of the daughter particles, the focus on the transverse momentum is motivated. Known effects such
 as sense-wire misalignment and particle energy loss induce a significant $p_T$-dependence to the momentum calibration, whereas the 
 polar-angle measurement is much less sensitive to $p_T$-dependent effects. Hence it is interesting to investigate the relationship 
 between $c^{\rm measured}$ and $c$ in the $p_T$-regime relevant to precision observables like $m_W$. In particular, we are 
 interested in the deviations of this relationship from exact equality. 

Since the invention of the multiwire proportional chamber~\cite{Charpak:1968kd}, this device and its derivative, the drift chamber, have 
 been used in a  myriad of experiments in particle physics and more broadly  in medicine, biology and radiation detection. 
 Their capabilities in terms of precision, accuracy, particle rate and radiation tolerance have steadily increased. In conjunction
 with a magnet, these tracking devices have been used to measure particle momenta with high precision and accuracy, such as 
 muons at 470~GeV by the E665 experiment~\cite{E665:1996mob,Kotwal:1995tu}. Most recently, the CDF Collaboraton has published an $m_W$ measurement  
 using the COT-based magnetic tracker whose particle-momentum measurement has been calibrated to 25 parts per million 
 (ppm)~\cite{CDF2022}. These
 and future precision momentum measurements with magnetic trackers motivate the analysis of these devices from first principles. 
 This paper presents a framework for understanding the COT momentum calibration, that may also be applicable to the precision
 magnetic trackers of the experiments at the LHC~\cite{atlasHiggs,cmsHiggs,lhcb}, future 
 colliders~\cite{ilc,cepc,clic,blondelFCCee,azzuriFCC}, and fixed-target experiments~\cite{moller,solid,p2}.  
\subsection{Analytic curvature response function}
\label{sec:analytic}
\hspace*{0.06in}
 With no loss of generality, an analytic response function with spatial degrees of freedom can be written as a Taylor expansion 
 around $c= 0$, the natural value of curvature in the absence of a magnetic field, 
$$ c^{\rm measured} = a_0 + (1 + a_1 + b_1q)c + (a_2 + b_2q)c^2 + (a_3 + b_3q)c^3 + ... $$
\noindent
where quartic and higher-order terms are not shown. It will be shown that, in the relevant range of $p_T$, 
  uncertainties due the higher-order terms are encapsulated  in the constraints on the parameters  up to the cubic coefficients. 
 This justifies the truncation of the Maclaurin expansion by information criteria~\cite{akaike}.

   The correspondence between the nomenclature used in this document and in previous CDF publications is described in Appendix~\ref{appendixModelCorrespondence}. 

The $c \to 0$ limit corresponds to the straight-line trajectory of a charged particle with very high $p_T$ when traversing an axial 
 magnetic field. From the perspective of a tracking device, $c = 0$ is just as natural as the situation with zero magnetic field, when 
 all charged-particle trajectories are straight lines. For this reason, the Maclaurin expansion above can equally well
  be considered as an expansion
 in the axial ($z$) component of the magnetic field ($B$), for a charged particle with a given $p_T$. In other words, the Maclaurin 
 expansion can be re-interpreted as an expansion in $B_z$. In a thought experiment, one can dial $B_z$ from negative to 
 positive values and the curvature of the particle's trajectory (of fixed $p_T$) will respond proportionately to $B_z$. Since $B_z$ 
 can be varied continuously and smoothly, the curvature and its measurement must track this smooth variation. 
 
 A perfect spatial 
 response, $c^{\rm measured} = c$,  implies that all $a$ and $b$ coefficients are zero. Therefore, 
$$ \delta c \equiv c^{\rm measured} - c = a_0 + (a_1 + b_1q)c + (a_2 + b_2q)c^2 + (a_3 + b_3q)c^3 + ... $$
where $c$ is defined at the beam axis. The latter coincides with the cylindrical axis of the COT, because the COT has been aligned with 
 the beam axis for all running conditions using  tracks of particles promptly-produced in beam-beam 
 collisions~\cite{cosmicAlignment,CDF2022}. The procedure for matching the COT axis and the beam axis is described in Appendix~\ref{appendixBeamCentering}. 
 The alignment between the COT and solenoid axes is discussed in Appendix~\ref{appendixSolenoid}.

\subsubsection{Charge dependence}
\hspace*{0.06in}
The presence of the two types of coefficients in the Maclaurin expansion allows for separate charge-independent  and charge-dependent  imperfections that
depend on the spatial (but not temporal) trajectory of the particle. This ansatz generalizes the model presented in~\cite{CDF2firstPRD}.

 The $a_0$ term represents the false curvature for a straight-line 
 trajectory which is typically induced by misalignment of the tracker sensors. The $a_1$ coefficient 
 represents the deviation from unity of the proportional momentum calibration factor, often referred to as the momentum scale factor or simply the momentum scale, and may be caused by the use of the incorrect value of the magnetic field or the tracker radius. The $b_2$ 
 coefficient is mimicked by the ionization energy loss incurred upstream of the tracker. These terms represent charge-symmetric effects. 

 The drift cells are tilted to compensate
  for the Lorentz angle  of the drifting electrons. The $b_1$, $a_2$ and $b_3$ coefficients  capture effects that might break the 
  charge symmetry due to this tilt. These charge-antisymmetric coefficients are shown to be negligible or have no impact on the CDF $m_W$ measurement~\cite{CDF2022}.

\begin{figure}[t]
\begin{center}
\includegraphics[width=3.2in]{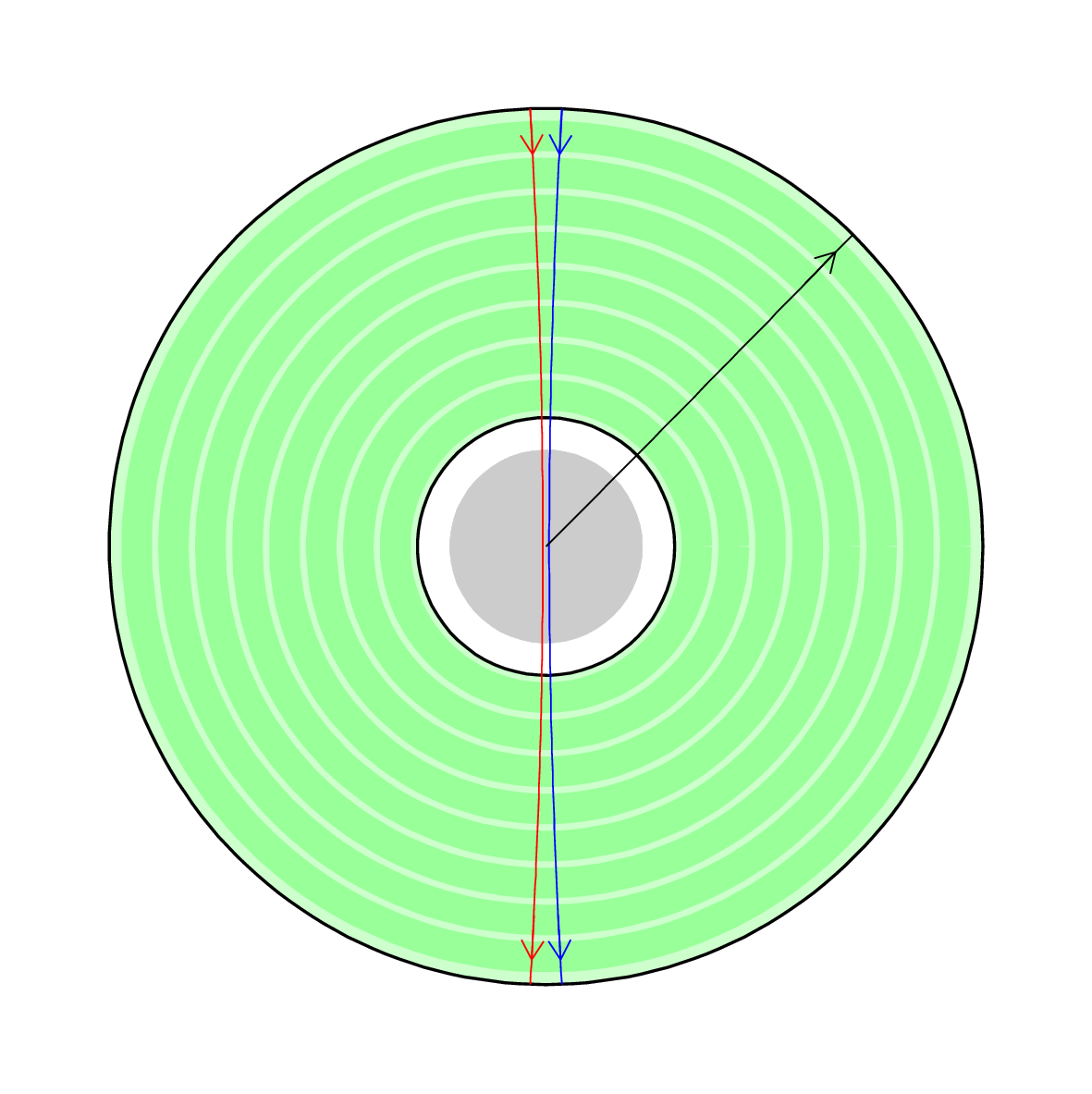}
\includegraphics[width=3.2in]{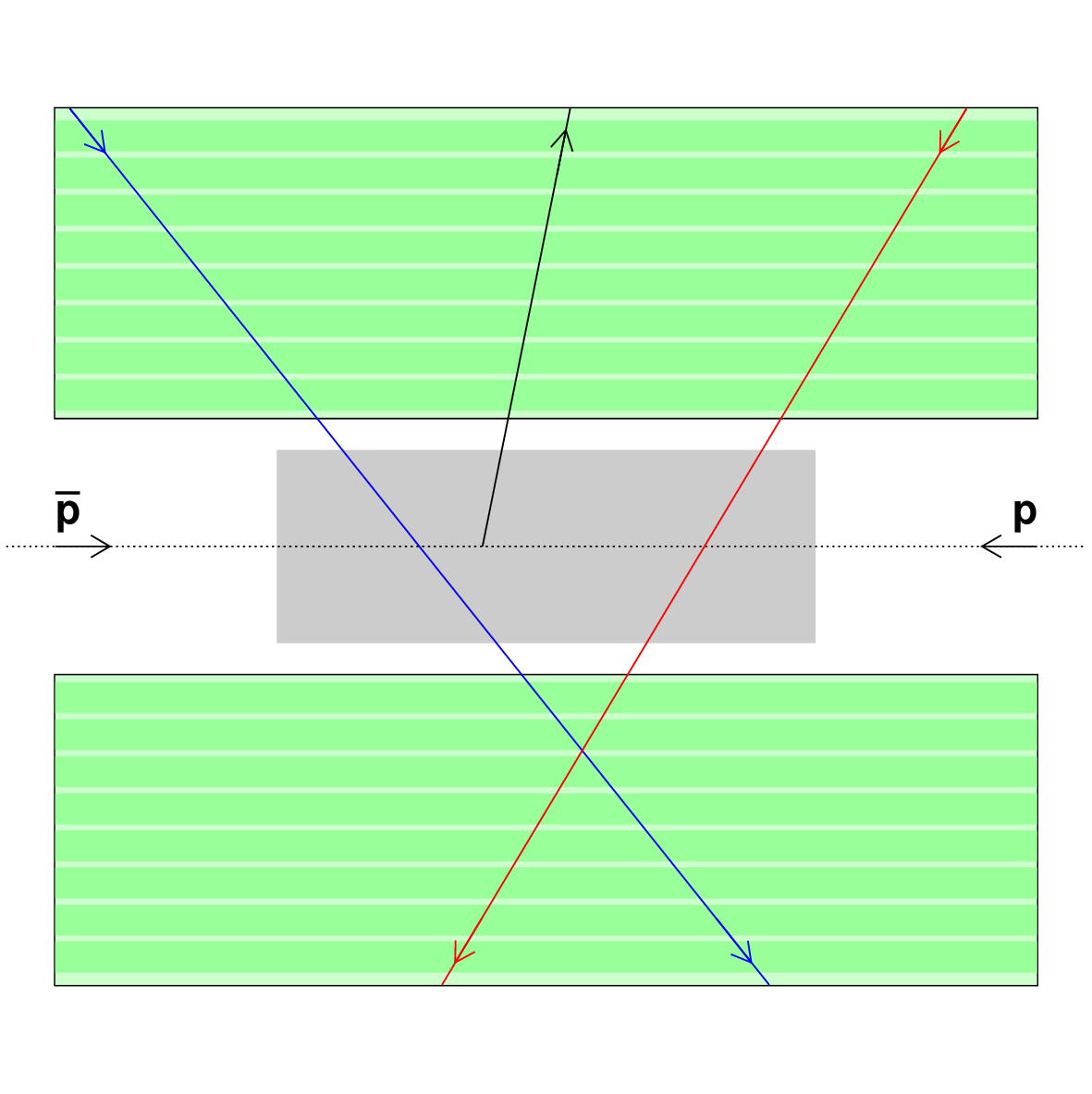}
\end{center}
\caption{Schematics of the CDF II experiment's cylindrical drift chamber (COT) in the transverse (azimuthal) view (left) and 
 longitudinal view (right). The beam line  is
 along the longitudinal axis of the tracker, which is also the direction of the magnetic field. The 
 green-shaded region shows the active gas-filled volume, between radii of 40~cm and 138~cm from the cylinder (and beam) 
 axis~\cite{cotNim}. The eight darker-green annuli depict the superlayers (rings of supercells) as shown in Fig.~\ref{fig:COTquad}.
 The grey shading indicates the region occupied by the silicon tracking detector. 
 The red and blue trajectories illustrate two oppositely charged cosmic-ray muons propagating downward in azimuth, each of $p_T = 
 10$~GeV. Their impact parameter of $\pm 1$~cm is typical of the cosmic-ray sample analyzed in this paper and in~\cite{cosmicAlignment}. 
 Also shown as the black trajectory is a muon with $p_T = 40$~GeV emanating from a decaying $W$ boson. 
 The arrows indicate the direction of propagation. The complete cosmic-ray trajectory is referred to as the dicosmic because it 
 is comprised of the incoming (pointing towards the beam axis) and outgoing (pointing away from the beam axis) legs of the 
 cosmic-ray path.  }
\label{fig:showCosmic}
\end{figure}
\subsubsection{Energy loss}
\hspace*{0.06in}
The energy $\epsilon$ lost by the charged particle as it traverses the tracker correlates with the temporal history of the particle. 
 By convention, $\epsilon > 0$ when the particle traverses the tracker in the outward radial direction, i.e. a particle produced in a  
 beam-beam collision or an outward-going cosmic ray (Fig.~\ref{fig:showCosmic}). 
 For these particles, $p_T^{\rm measured} = (p_T - \epsilon)$ since 
 the COT measurement occurs after the particle (produced with $p_T$) traverses the beam pipe and the 
 silicon detector which are situated upstream of the COT. The energy loss almost entirely occurs in these upstream devices and support
 structures. As a correction to $p_T$, $\epsilon$ is defined at normal incidence. 

Figure~\ref{fig:showCosmic} also shows that for an incoming cosmic ray, $p_T^{\rm measured} = (p_T + \epsilon)$ since the COT measurement occurs before the particle loses energy
 in the material between the COT and the beam axis (where $p_T$ is defined). To account for the bi-directional nature of cosmic-ray tracks,
 we introduce a binary variable $t$, with $t \equiv \texttt{+}1 (\texttt{-}1)$ for an outgoing (incoming) cosmic-ray track; 
 $t = \texttt{+}1$ for particles produced  in beam-beam collisions (Fig.~\ref{fig:showCosmic}). 
 
In Appendix~\ref{appendixEloss} it is shown that the energy loss induces, up to third order, the corrections terms $(tq \epsilon c^2 + \epsilon^2 c^3)$ to the equation above, 
\begin{equation}
\delta c = a_0 + (a_1 + b_1q)c + (a_2 + [b_2 + t \epsilon]q)c^2 + ([a_3 + \epsilon^2] + b_3q)c^3 + ... 
\label{eq:analytic}
\end{equation}                                                                                                                             
 Thus, the energy loss induces a $b_2$-like coefficient which is distinguishable from the spatial (geometry-induced) $b_2$ coefficient
 by comparing incoming $(t \equiv \texttt{-}1)$ and outgoing $(t \equiv \texttt{+}1)$ cosmic-ray tracks. 
 The energy loss also induces an $a_3$-like coefficient
 at $2^{\rm nd}$ order which is indistinguishable from geometrical sources of $a_3$.  

Hard scattering from the drift chamber wires is a negligible effect, as discussed in Appendix~\ref{appendixEloss}.
\subsubsection{Spatial uniformity}
\hspace*{0.06in}
 The coefficients $a_n$, $b_n$ and $\epsilon$ of the response function may initially depend on the azimuthal and polar angle at which the particle 
 traverses the drift chamber. Averaged  over $c$, these dependences have been eliminated by an alignment procedure\footnote{This average was 
 performed such that $\langle c \rangle \equiv 0$ for the alignment-data sample~\cite{cosmicAlignment}.} 
 that makes the COT response uniform with respect to orientation~\cite{cosmicAlignment}. 
 A summary of the spatial uniformity of the COT is provided in Appendix~\ref{appendixUniformity}.

 We will show that by far the largest sources of systematic uncertainty in the momentum calibration of the COT are the $a_1$ and $\epsilon$ parameters.
 
 The curvature proportionality parameter $a_1$ depends on the COT radius and the magnetic field. The COT radius is defined by its end plates and
 cannot depend on polar angle. The wires deviate from a straight-line shape due to gravitational and electrotatic deflections. A detailed analysis 
 and corrections for these deflections are presented in~\cite{cosmicAlignment} and summarized in Appendix~\ref{appendixUniformity}. A more
 significant source of polar-angle dependence of $a_1$ is the non-uniformity of the magnetic field due to the fringe-field effect at the edges
  of the solenoid. This effect is measured and corrected for using the $J/\psi \to \mu \mu$ data~\cite{CDF2firstPRD,CDF2014,CDF2022}.
 
 The energy loss $\epsilon$ is proportional 
 to the length of the path traversed; for the high-$p_T$ particles used in the $m_W$ analysis~\cite{CDF2022}, the dependence of the path 
 length on the curvature is negligible and the latter scales solely as $(\csc \theta)$ where $\theta$ is the polar angle in cylindrical 
 coordinates. As $\epsilon$ is defined at normal incidence, its impact on the measured curvature is nearly independent of $\theta$. While 
 the distribution of material along the beam axis is not perfectly uniform due to the placement of silicon-detector bulkheads, the 
 dependence of $\epsilon$ on this cylindrical $z$-coordinate is averaged over in a sufficiently similar way by all data samples used 
 in the $m_W$ analysis. The tracking detectors are also fairly symmetric in azimuth~\cite{svxII,cotNim}.

 Thus the salient features of the COT response  are captured by an inclusive study of the $a_n$, $b_n$ and $\epsilon$ coefficients. A
  quantitative assessment of this conclusion is provided in Appendix~\ref{appendixUniformity}.
\subsection{Non-analytic response}
\label{sec:nonanalytic}
\hspace*{0.06in}
It is conceivable that gaps in the acceptance of a tracking detector due to uninstrumented or dead regions, or boundaries between sensors lead to 
non-analytic behavior of the response function. An investigation of this possibility is presented in Sec.~\ref{sec:singular} in the form of
$(a_r + b_r q)|c|^r$ terms with negative or fractional values of $r$. We find that the worst-case scenario is adequately parameterized by a
$b_0 q$ term; other terms are either redundant or inconsistent with the COT being a single, unfragmented tracking volume. Thus, Eq.~\ref{eq:analytic}
may be fully generalized by including this term. In Sec.~\ref{sec:singular}
 we discuss constraints on $b_0$ from studies of COT hit efficiency and drift displacement in the $c \to 0^\pm$ limit. 
\section{Constraints from cosmic-ray tracks}
\label{sec:cosmics}
\hspace*{0.06in}
Cosmic rays provide a powerful control sample of data that CDF has used to pin down many attributes of the drift chamber. There are two salient
features of this sample. First, high-$p_T$ cosmic-ray muons are automatically selected by the same trigger paths that acquire  the $W$ and $Z$ boson
data in the muon channel. This feature guarantees that the cosmic-ray events are chosen in situ with collider data and that they experience the
same operating conditions of the detector as the collider data used for physics analysis. The trigger-timing requirements ensure that these cosmic
rays are synchronous with proton-antiproton collisions within a few nanoseconds~\cite{cotcosmic}. In addition to the five spatial parameters that characterize 
tracks originating from $p \bar{p}$ collisions, cosmic-ray tracks include the beamline-crossing time and the direction of propagation
as fitted parameters. These track-fitting procedures ensure the equivalence of the fitted spatial parameters between the two types of tracks~\cite{cotcosmic}.
For these reasons, measurements of the drift chamber made with  cosmic rays are usable for reconstructed tracks in collider data. 

Second, cosmic rays traversing the COT through all its radial layers, and passing close to the beam axis, contain a very useful redundancy that
CDF exploits for accurate calibration. 
 Figure~\ref{fig:showCosmic} shows that cosmic rays provide an excellent data sample for tracker alignment and bias measurement~\cite{cosmicAlignment}. The comparison of the two
 legs of the reconstructed cosmic-ray trajectory provides an estimate of track parameter biases because the two 
 legs provide independent measurements of the same particle (up to the muon's ionization energy loss as 
 it traverses the silicon tracker)~\cite{cosmicAlignment}. The distribution of the cosmic rays is fairly uniform in azimuth and in 
 polar angle.  

 The measured curvature of the outgoing leg is  
$$ c^{\rm out} = a_0 + (1 + a_1 + b_1q)c + (a_2 + [b_2 + \epsilon]q)c^2 + (a_3 + \epsilon^2 + b_3q)c^3 + ... $$
and the incoming leg is reconstructed as a muon of the opposite charge, $c \to -c$ and $q \to -q$, which is also time-reversed, $t \to -t$, 
\begin{eqnarray} 
c^{\rm in} & = & a_0 + (1 + a_1 - b_1q)(-c) + (a_2 - [b_2 - \epsilon]q)(-c)^2 + (a_3 + \epsilon^2 - b_3q)(-c)^3 + ... \nonumber \\ 
           & = & a_0 + (-1 - a_1 + b_1q)c + (a_2 - b_2q + \epsilon q)c^2 + (-a_3 - \epsilon^2 + b_3q)c^3 + ... \nonumber
\end{eqnarray}
Therefore, 
\begin{equation}
\Delta^+_c = \frac{1}{2}(c^{\rm out} + c^{\rm in}) = a_0 + b_1qc + (a_2 + \epsilon q)c^2 + b_3qc^3 + ...
\label{deltaPlusEquation}
\end{equation}
 and 
\begin{equation}                                                                                                                     
    \Delta^-_c = \frac{1}{2}(c^{\rm out} - c^{\rm in}) = (1 + a_1)c + b_2qc^2 + (a_3 + \epsilon^2)c^3 + ...
\label{deltaMinusEquation}             
 \end{equation}

For the right-hand-side of these equations, the true value of $c$ is well-represented by the measurement of the ``dicosmic helix'', which
 is described in~\cite{cosmicAlignment} and~\cite{cotcosmic}. The dicosmic helix fit is performed on the combined set of COT hits 
 associated with the incoming and outgoing legs together (Fig.~\ref{fig:showCosmic}). The direction of the momentum of the cosmic-ray muon is taken into account 
 in this fit (and in the separate fits to the incoming and outgoing legs for the measurement of $c^{\rm in}$ and $c^{\rm out}$  respectively)
 by adjusting each drift chamber hit coordinate for the appropriate time-of-flight delay~\cite{cotcosmic}. The same time-of-flight delay  is 
 included in the helix fit to all outgoing drift-chamber tracks of particles emanating from beam-beam collisions that are 
 used for physics analysis. 

 The curvature measurement $c_{\rm d}$ from the dicosmic helix fit is more precise than the measurements of $c^{\rm in}$ and $c^{\rm out}$ by a factor 
 of $8 \sqrt{2}$, due to the track length increase from 96~cm to 274~cm and the doubling of 
 the number of hits. Hence the dicosmic helix measurement provides a good
 proxy for the true $c$.  The corollary is that ${\rm sgn}(c_{\rm d})$ is a good proxy for $q$ as long as we avoid sensitivity 
  to resolution-induced bias. The resolution of $c^{\rm in}$ and $c^{\rm out}$ is measured to be 
 1.3~TeV$^{-1}$~\cite{cosmicAlignment}, therefore the resolution of $c_{\rm d}$ is expected to be 110~PeV$^{-1}$.   

The cosmic-ray data presented here are identical to the data presented in~\cite{cosmicAlignment}. The track impact parameter $(d_0)$ 
 with respect to the beam axis is distributed such that  $|d_0| < 3$~cm and the $z$-coordinate of the tracks' point of closest approach to 
 the beam axis $(z_0)$ is required to be within 60~cm of the beam-beam collision point. These conditions ensure that the cosmic-ray tracks 
 have similar trajectories as the particles selected for the $m_W$ measurement~\cite{CDF2022}.   
\subsection{Constraints from $\Delta^+_c$}
\begin{figure}
\begin{center}
\includegraphics[width=3.5in]{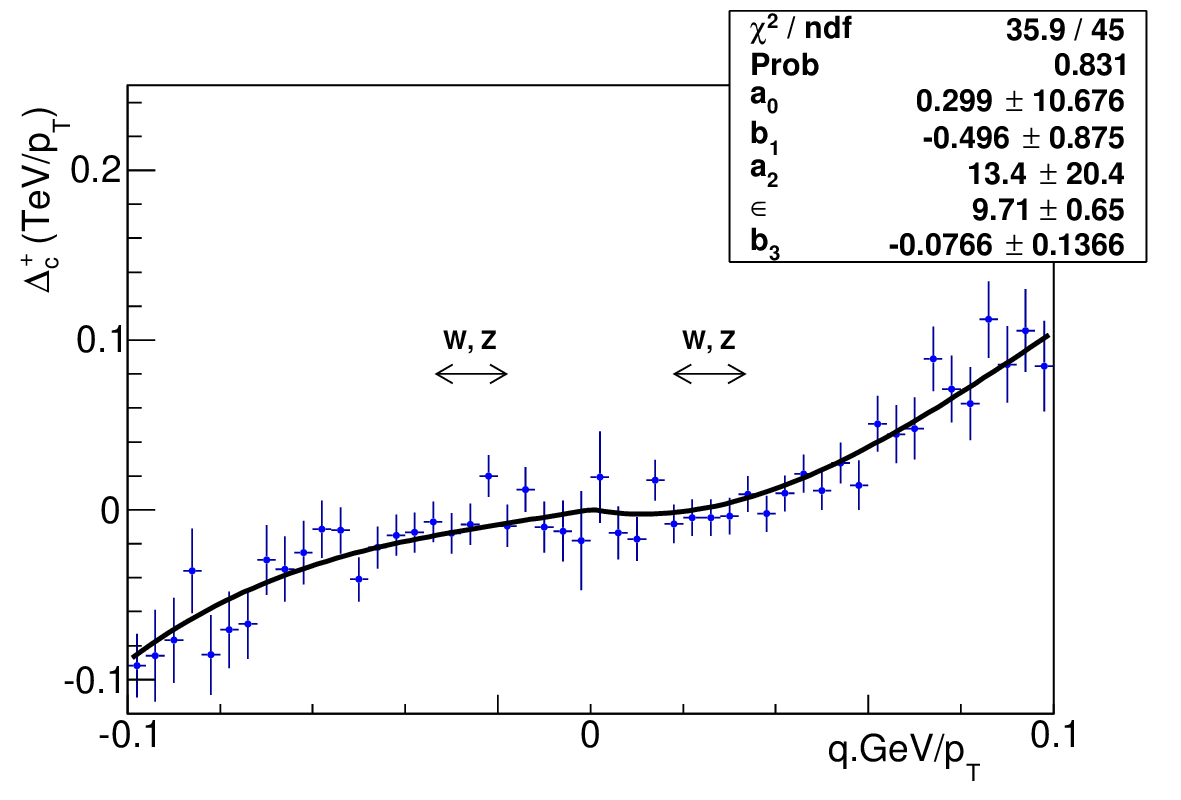}
\includegraphics[width=3.5in]{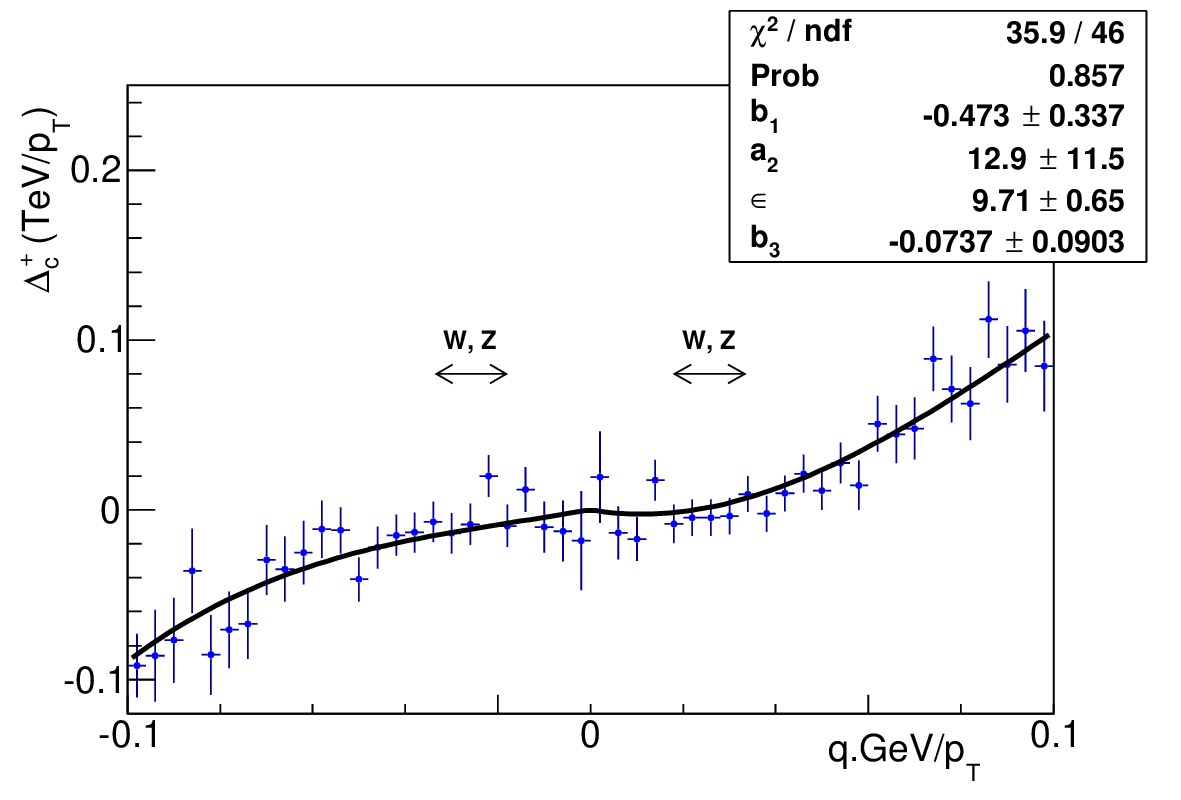}
\end{center}
\caption{The measurement of $\Delta^+_c$ as a function of $c_{\rm d}$, the measured curvature of the COT dicosmic helix, in cosmic-ray data collected in situ during collider operation. The requirement $| z_0 | < 60$~cm ensures that the cosmic-ray tracks 
 have similar trajectories as the particles selected for physics analysis. 
 Also shown are the fits to Eq.~\ref{deltaPlusEquation} and the values and statistical uncertainties 
 of the fitted parameters (left) $a_0$ (in 
 PeV$^{-1}$), $b_1$ (in \permil), $a_2$ and $\epsilon$ (both in MeV), and $b_3$ (in GeV$^2$), and (right) $b_1$, $a_2$ and $\epsilon$, 
 and $b_3$. The error bars indicate the statistical uncertainties on the data points. The horizontal arrows indicate the range of
 $q/p_T$ of the leptons originating from $W^\pm  \to \ell^\pm \nu$ and $Z \to \ell^+ \ell^-$ decays that are used in the $m_W$ 
 analysis~\cite{CDF2022}.}
\label{deltaPlus}
\end{figure}

\hspace*{0.06in}
The measurement of $\Delta^+_c$ as a function of $c_{\rm d}$ is shown in Fig.~\ref{deltaPlus}. The 5-parameter fit of Eq.~\ref{deltaPlusEquation} 
 to the data is superposed, 
 along with the fitted values of the parameters. Since these data are post-alignment~\cite{cosmicAlignment}, it is expected that 
 the fitted value of $a_0$ is statistically consistent with zero. The remaining parameters provide new information about the COT
 response. As shown in Appendix~\ref{appendixBplus}, this analysis is expected to provide accurate measurements of the coefficients
 in Eq.~\ref{deltaPlusEquation}. 

The fit to these cosmic-ray data finds a value for the energy loss $\epsilon = (9.71 \pm 0.65_{\rm stat})$~MeV. This measurement is 
 consistent with the value\footnote{The first-principles value of $\epsilon$ is     
 obtained in~\cite{CDF2firstPRD,CDF2022} by {\it ab initio} calculation of the  energy loss, based on published formulae and a fine-grained
    3-dimensional lookup table of the amount of ionizing material used during construction of the silicon tracker and the COT.}
  quoted in Sec.~V~B3 of~\cite{CDF2firstPRD}, ``Each muon passing through the silicon and COT detectors
loses on average 9 MeV at normal incidence''. 
As the $\epsilon$ term is the only term odd in $q$ in Eq.~\ref{deltaPlusEquation}, 
 $\epsilon$ is minimally correlated with the other parameters; the largest correlation 
 coefficient between $\epsilon$ and any other parameter is 8\%. 

Since the alignment procedure~\cite{cosmicAlignment} brings $a_0$ close to zero, it is a redundant parameter in the fit shown in 
 Fig.~\ref{deltaPlus} (left). The fit is repeated after constraining $a_0 = 0$ in order to obtain the approriate constraints on 
 the parameters $b_1$, $a_2$ and $b_3$. As shown in Fig.~\ref{deltaPlus} (right),  the latter
  are found to be statistically consistent with zero, implying the absence of these imperfections
 in the COT within the respective precisions. The uncertainties on these parameters will be used to bound the corresponding
 uncertainties in the $m_W$ measurement in Sec.~\ref{sec:mass}.  In this fit, $a_2$ is strongly anticorrelated with $b_1$ and 
 $b_3$, with correlation coefficients of $-97$\% and $-98$\% respectively.    
\subsection{Comment on $\Delta^-_c$}
\label{sec:deltaCm}
\hspace*{0.06in}
The measurement of $\Delta^-_c$ as a function of $c$, if the true $c$ were known, would provide information on the coefficients in
 Eq.~\ref{deltaMinusEquation}. However, if  $c_{\rm d}$ is used as a proxy for $c$, no information can be extracted from $\Delta^-_c$, 
 as shown in Appendix~\ref{appendixBminus}. 
\section{Constraint from positron-electron difference of $\langle E/p \rangle$ }
\label{sec:eop}
\hspace*{0.06in}
The calorimeter measurement of the energy $E$ of electrons and positrons is combined with the polar-angle measurement of the track to obtain
 their transverse energy $E_T$. For electrons and positrons of the same momentum incident at the same location of the calorimeter, 
 their electromagnetic shower energies are identical within (1 MeV)/$E \sim$~(1~MeV)/(40~GeV)~$\lesssim 30$~ppm, 
 where 40~GeV is the 
 relevant $E_T$ for $W$ and $Z$ boson decays and 1~MeV is the additional energy released from the positron annihilation. Hence the 
 difference of $\langle E_T/p_T \rangle$ between positrons and electrons is used to constrain the tracker bias in the measurement of 
 $p_T$. 
  
Using the calibrated measurement of $E_T$ as a proxy for $q/c$, we write this difference as
\begin{eqnarray}
\Delta_{pe} & \equiv & \frac{1}{2}[\langle \frac{E_T}{p_T}(e^+)\rangle - \langle \frac{E_T}{p_T}(e^-) \rangle] 
= \frac{1}{2}\Sigma_q q \langle \frac{E_T}{p_T}\rangle = \frac{1}{2}\Sigma_q \frac{q}{c}c^{\rm measured}  
= \frac{1}{2}\Sigma_q \frac{q}{c}(c + \delta c) \nonumber \\
& = & \frac{1}{2}\Sigma_q q \frac{\delta c}{c}
= \frac{1}{2}\Sigma_q \frac{q}{c} [a_0 + (a_1 + b_1q)c + (a_2 + b_2^{\prime}q)c^2 + (a_3 + \epsilon^2 + b_3q)c^3 ] \nonumber \\ 
& = & \frac{1}{2}\Sigma_q [a_0\frac{q}{c} + (a_1q + b_1) + (a_2q + b_2^{\prime})c + (a_3q + \epsilon^2q + b_3)c^2] \nonumber \\ 
& = &  a_0 \langle p_T \rangle + b_1 + a_2 \langle p_T^{-1} \rangle + b_3 \langle p_T^{-2} \rangle
\label{eq:deltaPE}
\end{eqnarray}
\noindent
since $a_1\Sigma_q q$, $b_2^{\prime}\Sigma_q c$ and $\Sigma_q(a_3 + \epsilon^2)qc^2$ vanish by charge symmetry. 
The combination of the two coefficients $b_2^\prime \equiv b_2 + \epsilon$ is motivated in Sec.~\ref{sec:mass1}.
 
In the $m_W$ analysis~\cite{CDF2022}, tracks are calibrated to eliminate this difference so that $\Delta_{pe} = 0 \pm 43$~ppm.  The statistical uncertainty on $\Delta_{pe}$ is the same\footnote{Taking the difference between two half-samples increases the 
 statistical uncertainty by a factor of $\sqrt{2} \times \sqrt{2}$ which is cancelled by the factor of $\frac{1}{2}$ in the definition
 of $\Delta_{pe}$.} as the statistical uncertainty on the fit to the inclusive  $E_T/p_T$  distribution shown in~\cite{CDF2022}. The bias 
 on $\Delta_{pe}$ due to the positron annihilation energy of $\lesssim 15$~ppm is negligible compared to the statistical uncertainty. 

As this method calibrates a quantity linear in $\delta c$, there is no need to consider
higher powers of $\delta c$.
\subsection{Bremsstrahlung}
\hspace*{0.06in}
While muons experience ionization energy loss as they traverse the inner silicon vertex detector, electrons/positrons
   also undergo energy loss due to 
  bremsstrahlung photon radiation. The radiated photons are almost always coalesced with the calorimeter shower produced by the primary $e^\pm$. In these radiative cases, 
  $E_T$ ($p_T$) measures the primary $e^\pm$ before (after) the bremsstrahlung emission and the spectrum of $E_T/p_T$ has a high-side radiative tail
  (e.g. Fig.~2B of~\cite{CDF2022}). Non-radiative $e^\pm$ candidates 
  are selected for the $\Delta_{pe}$ measurement by requiring $0.9 < E_T/p_T < 1.1$~\cite{CDF2022,CDF2014}. The interval isolates the resolution-broadened peak of
  $E_T/p_T$ near unity.  The soft bremsstrahlung within this interval acts as a 
 percent-level scale factor on $\Delta_{pe}$ which is irrelevant when the alignment brings $\Delta_{pe}$ to zero.
\section{Bias in invariant-mass fits}
\label{sec:mass}
\hspace*{0.06in}
The reconstruction of the invariant mass $m$ 
 of a neutral particle from its two-body decay into
 massless particles can
 be written as follows:
$$m^2 = (p_1 + p_2)^2 = 2 p_1.p_2 = 2 E_1E_2(1-\cos\gamma) = 2 p_{T1}p_{T2}(1-\cos\gamma)/
(\sin\theta_1 \sin\theta_2)$$
\noindent
where $\gamma$ is the opening angle between the 3-vectors of the daughters and $\theta_{1,2}$
 are their respective polar angles. Since the uncertainty on the angle measurements has a negligible contribution to the mass uncertainty 
 in comparison to the curvature uncertainty~\footnote{The uncertainty due to the polar angle is 4~ppm, 8~ppm and 11~ppm respectively for the 
 $J/\psi$, $\Upsilon$ and $Z$ boson mass measurement~\cite{CDF2022}. These uncertainties have minimal impact given the total  uncertainties of
 29~ppm, 36~ppm and 70~ppm on the respective momentum calibrations derived from these data~\cite{CDF2022}. The azimuthal angle is measured $5\times$($30\times$) more
 accurately than the polar angle~\cite{cosmicAlignment} without (with) the beam constraint, and does not contribute an uncertainty on the mass measurements.},
$$m^2 \propto 2p_{T1}p_{T2} = 2\frac{q_{1}}{c_{1}}\frac{q_{2}}{c_{2}} =  \frac{-2}{c_{1}c_{2}}$$ 
\subsection{First-order effects on mass reconstruction}
\label{sec:mass1}
\hspace*{0.06in}
The bias in the measured mass, $\delta m$ caused by the COT biases is
$$m \delta m |_{\rm 1st} \propto \frac{1}{c_2}\frac{\delta c_1}{c_1^2} + \frac{1}{c_1}\frac{\delta c_2}{c_2^2}$$ 
\noindent
at first-order and the corresponding fractional mass bias is
$$\frac{\delta m}{ m } |_{\rm 1st} =  
(\frac{1}{c_2}\frac{\delta c_1}{c_1^2} + \frac{1}{c_1}\frac{\delta c_2}{c_2^2})/m^2 
 = 
(\frac{1}{c_2}\frac{\delta c_1}{c_1^2} + \frac{1}{c_1}\frac{\delta c_2}{c_2^2})\frac{c_1c_2}{-2}
= 
\frac{-1}{2}(\frac{\delta c_1}{c_1} + \frac{\delta c_2}{c_2})      
 $$

Using the response model of Eq.~\ref{eq:analytic}, 
 the fractional mass bias can be expressed in terms of the $a$, $b$ and $\epsilon$ coefficients. The first bias term is 
$-a_0 \langle c^{-1} \rangle  = -a_0 \langle q \cdot p_{T} \rangle$ which 
averages to zero since the decay is charge-symmetric. Therefore $a_0$ can only induce a 
 mass bias at quadratic order, and by dimensional analysis the fractional mass bias must 
 be proportional to $(a_0 p_T)^2$, as shown in Sec.~\ref{sec:mass2}. 

The next terms causing a fractional mass bias are $\frac{1}{2}\Sigma_q (a_1 + b_1q)$. The $b_1$ term 
cancels between the two opposite charges and is not observable in the invariant mass. The
 $a_1$ term is the momentum calibration correction and therefore the fractional mass bias is proportional to
 $a_1$. Thus, $a_1 = 0$ after the precise calibration based on the $J/\psi \to \mu \mu $
 and $\Upsilon \to \mu \mu $ mass fits and the corresponding uncertainty have been incorporated in the $m_W$ analysis~\cite{CDF2022}. 

 The next two terms causing a mass bias are $\frac{1}{2}\Sigma_q (a_2 + [b_2 +t \epsilon]q)c$. Again by the charge
 symmetry of the decay, $a_2 \Sigma_q c = 0$ so $a_2$ 
 is not visible in an inclusive mass fit\footnote{The $a_2$ term should be visible as the slope in a plot of $\delta m/m$ 
versus $\langle c \rangle$.}.

 As invariant mass fits are performed for outgoing particles only (emanating from beam-beam collisions), $t=1$ in this context (see 
 Sec.~\ref{sec:intro}) and we can use the combined $b_2 + \epsilon = b_2^\prime$ for this discussion.  
 The term $\frac{1}{2}b_2^\prime \Sigma_q qc = \frac{1}{2}b_2^\prime \Sigma_q p_T^{-1} = b_2^\prime \langle p_T^{-1} \rangle$.
 Thus the slope of $\delta m/m$ versus $\langle p_T^{-1} \rangle$ measures $b_2^\prime$. The 
 $b_2^\prime$ coefficient  is tuned from the $J/\psi \to \mu \mu$ mass fits in bins of 
 $\langle p_T^{-1} \rangle$ with an uncertainty of 34~keV in the $m_W$ analysis~\cite{CDF2022}. Thus, we can set $b_2^\prime = 0$
 after this calibration. The corresponding uncertainties at the three mass scales is,
\begin{itemize}
\item at $m_{W}$ and $m_{Z}$, $p_T \sim 40$~GeV: (34~keV)/(40~GeV)~$\sim 1$~ppm
\item at $m_{\Upsilon}$, $p_T \sim 5$~GeV: (34~keV)/(5~GeV)~$\sim 7$~ppm
\item at $m_{J/\psi}$, $p_T \sim 3.3$~GeV: (34~keV)/(3.3~GeV)~$\sim 10$~ppm
\end{itemize}
This uncertainty on $m_{W,Z}$ is negligible and the uncertainty on $m_{\Upsilon}$ and $m_{J/\psi}$ is included in the $m_W$ 
 analysis~\cite{CDF2022}.

The next two terms causing a mass bias are $\Sigma_q ([a_3 + \epsilon^2] + b_3q)c^2$. Again by the charge
 symmetry of the decay, $b_3 \Sigma_q qc^2 = 0$ so $b_3$
 is not visible in an inclusive mass fit. Therefore the surviving term is $[a_3 + \epsilon^2]\langle c^2 \rangle$. The order of magnitude 
 of the $\epsilon^2$ term at the three relevant mass scales is,
\begin{itemize}
\item at $m_{W}$ and $m_{Z}$, $p_T \sim 40$~GeV: $\epsilon^2 \langle c^2 \rangle \sim [(9 {\rm ~ MeV}) / (40 {\rm ~ GeV})]^2 \sim 0.1$~ppm
\item at $m_{\Upsilon}$, $p_T \sim 5$~GeV: $\epsilon^2 \langle c^2 \rangle \sim [(9 {\rm ~ MeV}) / (5 {\rm ~ GeV})]^2 \sim 3$~ppm
\item at $m_{J/\psi}$, $p_T \sim 3.3$~GeV: $\epsilon^2 \langle c^2 \rangle \sim [(9 {\rm ~ MeV}) / (3.3 {\rm ~ GeV})]^2 \sim 7$~ppm
\end{itemize}
Not only are these energy-loss effects small at  quadratic  order, they are already accounted for in the simulated line-shape templates
 used for these mass fits~\cite{CDF2022}, with vanishing uncertainty. 

 In principle, the $J/\psi \to \mu \mu$ data can be used to measure $a_3$ by fitting for a quadratic dependence of $\delta m/m$ versus 
 $\langle p_T^{-1} \rangle$. In practice, the linear fit to these data using $a_1$ and $b_2^\prime$ as free parameters is found to 
 be consistent with the data within their uncertainties~\cite{CDF2022}. By the Akaike Information Criterion (AIC)~\cite{akaike}, this implies that 
 the inclusion of $a_3$ as a fit parameter would find a statistically insignificant value, and that the $a_3$ uncertainty is already 
 accounted for when the uncertainties on $a_1$ and $b_2^\prime$ are propagated to low $p_T^{-1}$. This procedure is conservative because
 the effect of $a_3$ vanishes more rapidly as $p_T^{-1} \to 0$ than the effect of $b_2^\prime$; therefore absorbing the uncertainty due 
 to $a_3$ into the uncertainty due to $b_2^\prime$ overestimates the propagated $m_W$ uncertainty. 

 An independent upper bound on the uncertainty due to   
 $a_3$ is obtained by comparing the $J/\psi \to \mu \mu$ and $\Upsilon \to \mu \mu$ data, as discussed below. 

 We conclude that a mass measurement is biased at first order by the parameters $a_1$, $b_2^\prime$ and $a_3$, while the remaining 
 parameters $a_0$, $b_1$, $a_2$ and $b_3$ do not introduce a first-order bias. The uncertainties due to $a_1$, $b_2^\prime$ and $a_3$ are 
 already  accounted for in the $m_W$ analysis~\cite{CDF2022}. 
\subsection{Second-order effects on mass reconstruction}
\label{sec:mass2}
\hspace*{0.06in}
We evaluate the mass bias induced at second order by the curvature response function. Considering the second-order derivatives, 
$$m \delta m |_{\rm 2nd} \propto -2\frac{1}{c_2}\frac{(\delta c_1)^2}{c_1^3} -2\frac{1}{c_1}\frac{(\delta c_2)^2}{c_2^3} - 2\frac{\delta c_1}{c_1^2}\frac{\delta c_2}{c_2^2}$$
$$ \frac{\delta m}{m} |_{\rm 2nd} = c_1 c_2 (\frac{1}{c_2}\frac{(\delta c_1)^2}{c_1^3} + \frac{1}{c_1}\frac{(\delta c_2)^2}{c_2^3} + \frac{\delta c_1}{c_1^2}\frac{\delta c_2}{c_2^2}) = 
 (\frac{\delta c_1}{c_1})^2 + (\frac{\delta c_2}{c_2})^2 + \frac{\delta c_1}{c_1}\frac{\delta c_2}{c_2} $$
$$ = (\frac{\delta c_1}{c_1} + \frac{\delta c_2}{c_2})^2 - \frac{\delta c_1}{c_1}\frac{\delta c_2}{c_2} = 
(2 \frac{\delta m}{m}|_{\rm 1st})^2 - \Pi_q (A + qB) $$
\noindent
where
\begin{equation}
  A \equiv a_1 + b_2^{\prime}|c| + a_3c^2 = a_1 + b_2^{\prime}/p_T + a_3/p_T^2
  \label{eq:A}
\end{equation}  
 contains the terms of the response function that induce a first-order bias and 
\begin{equation}
  B \equiv q(a_0/c + b_1q + a_2c + b_3qc^2) = a_0p_T + b_1 + a_2/p_T + b_3/p_T^2
  \label{eq:B}
\end{equation}
  contains the terms that do not  induce a first-order bias. Note that $A$ and $B$ are symmetric in charge, and $\delta c/c \equiv A + qB$. 

In Sec.~\ref{sec:mass1} it was shown that the first-order fractional mass bias was equal to $A$. 
Thus, at second order,
 $$ \frac{\delta m}{m}|_{\rm 2nd} = (2A)^2 -(A+B)(A-B) = B^2 + 3A^2 \approx B^2 = \Pi_q (a_0p_{T} + b_1 + a_2p_{T}^{-1} + b_3p_T^{-2})$$
where the product is over the two (oppositely-charged) daughter particles. 
The square of the first-order bias $A^2 <(10^{-4})^2$ is negligible, therefore the second-order fractional mass bias is 
 approximated by $B^2$.   
\subsubsection{$2^{\rm nd}$-order effects at the $m_{W,Z}$ scale}
\label{sec:mWZ2}
Figure~\ref{deltaPlus} provides the following 
 estimates of the uncertainties due to each of the terms in the second-order fractional mass bias, at a typical $p_T \sim 40$~GeV, using 
 the expression for $B^2$.
\begin{itemize}
\item $a_0$ term: ($11 ~ {\rm PeV}^{-1} \cdot 40$~GeV)$^2 \sim$ (440~ppm)$^2 \sim 0.2$~ppm 
\item $b_1$ term: (0.5 \permil)$^2 \sim 0.3$~ppm
\item $a_2$ term: $(\frac{14 ~ {\rm MeV}}{40 ~ {\rm GeV}})^2 \sim 0.2$~ppm 
\item $b_3$ term: $[(0.09 ~ {\rm GeV}^2) (40~ {\rm GeV})^{-2}]^2 \sim (60~{\rm ppm})^2 \sim 3$~ppb (parts per billion) 
\end{itemize}
 These estimates show that, even ignoring the large anticorrelation between these coefficients as inferred from the fits in 
 Fig.~\ref{deltaPlus}, the second-order bias in $W$ and $Z$ boson mass fits is expected to be negligible.

 In the next two sections the effect of the large anticorrelation between these coefficients is estimated by removing redundant parameters in
 the fit to the cosmic-ray data (Fig.~\ref{deltaPlus}). We will find that the uncertainty on $a_2$ reduces by a factor of 5, thereby reducing
 its $2^{\rm nd}$-order effect at the $m_{W,Z}$ scale by a factor of 25 to 8~ppb. Thus, only $a_0$ and $b_1$ are relevant for estimating
 the $2^{\rm nd}$-order effects at the $m_{W,Z}$ scale. This is expected since $a_0$ and $b_1$ are low-curvature terms while $a_2$ and $b_3$ are
 high-curvature terms.

 For the same reason, the observables obtained from the $W \to \ell \nu$ data, $\Delta_{pe}$ (see Sec.~\ref{sec:eop}) and $\Delta_W^\mp$ (to be
  introduced in Sec.~\ref{sec:wmass1}) constrain $a_0$ and $b_1$ but not $a_2$ and $b_3$. 
  \subsubsection{$2^{\rm nd}$-order effects at the $m_{\Upsilon}$ scale}
  \label{sec:mUpsilon2}
Since $b_1$ is dimensionless, the insignificant second-order bias induced by this parameter is independent of curvature and the mass being 
 reconstructed. Hence it is a negligible parameter and can be removed from the response model. Similarly, as the energy loss $\epsilon$ 
 is measured precisely from the $J/\psi \to \mu \mu$ data and incorporated in the $b_2^\prime$ parameter, we can correct the cosmic-ray
 measurements for its known value (which is consistent with the fitted value in Fig.~\ref{deltaPlus}) and remove it as a free parameter. 

 The elimination of redundant parameters is equivalent to propagating the covariance matrix (including correlation coefficients that 
 approach $\texttt{-}1$ for redundant parameters) on the original, complete set of parameters. In practice, redundant parameters lead to 
 numerical instabilities; hence the removal of redundant parameters yields robust estimates. 

\begin{figure}
\begin{center}
\includegraphics[width=3.5in]{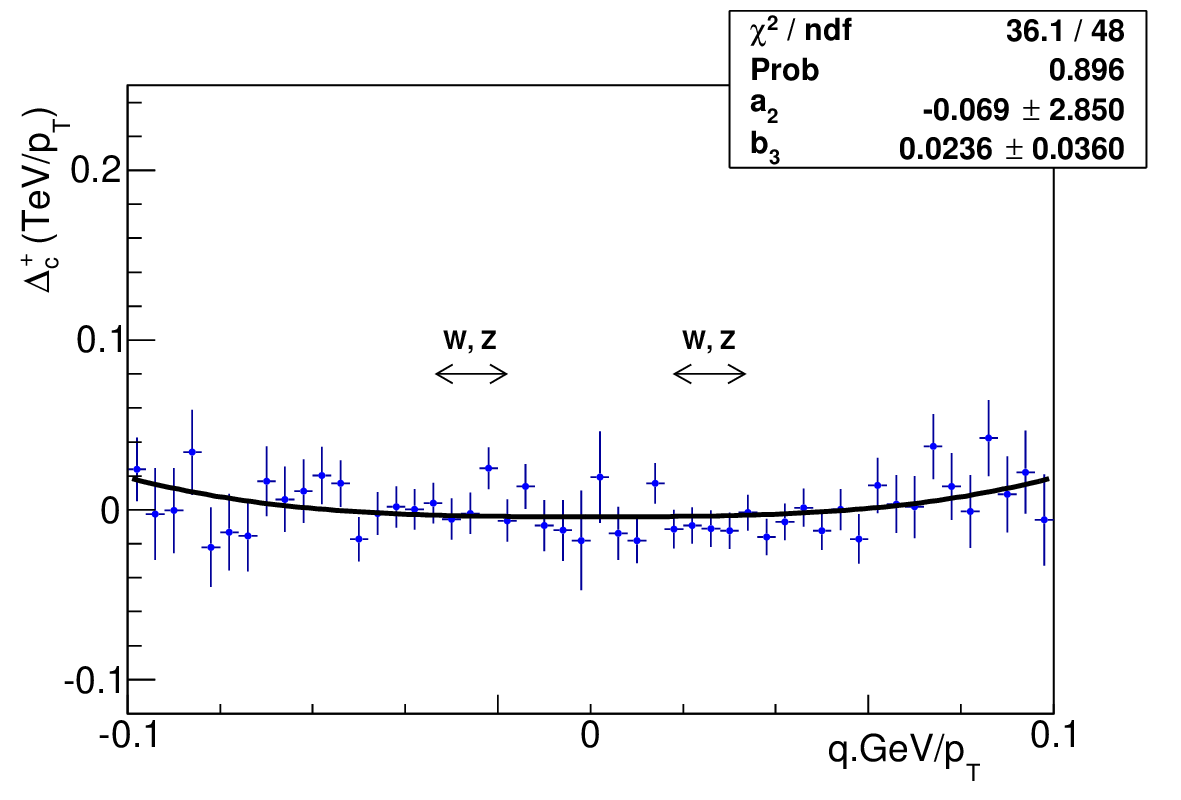}
\includegraphics[width=3.5in]{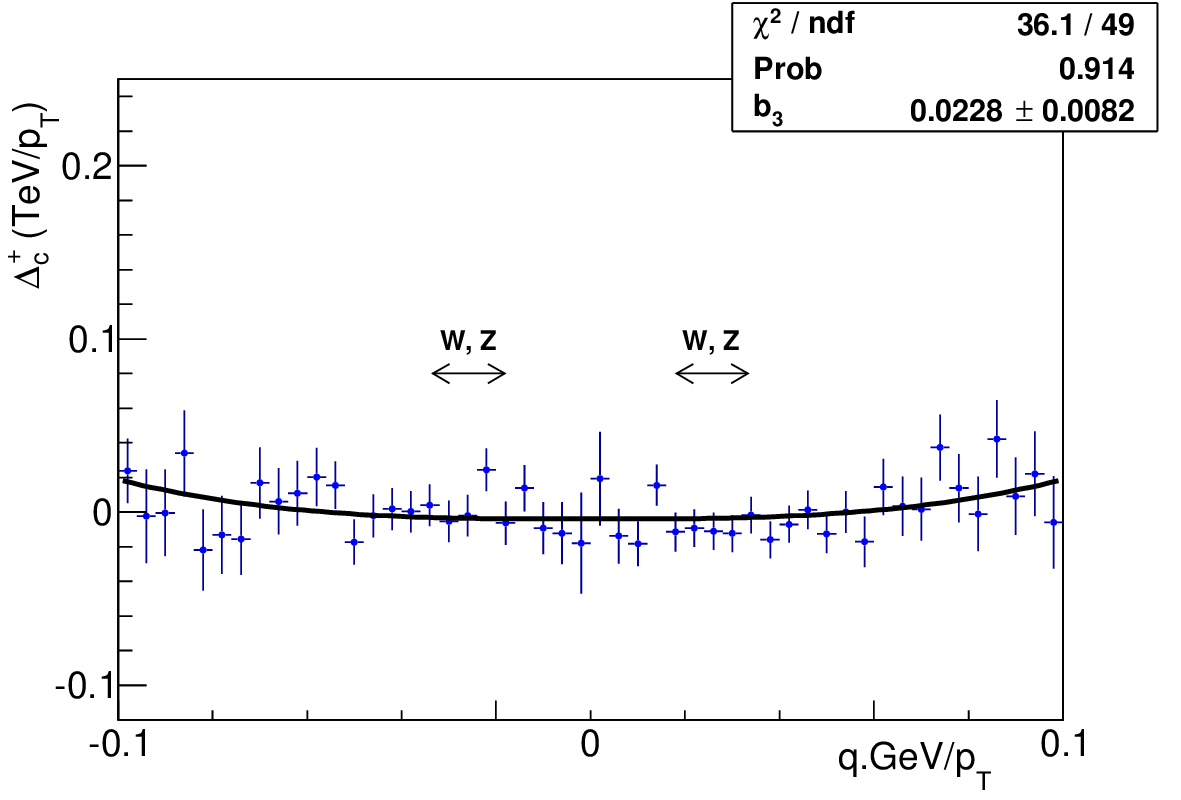}
\end{center}
\caption{The measurement of $\Delta^+_c$ as a function of $c_{\rm d}$, the measured curvature of the COT dicosmic helix, in cosmic-ray data collected in situ during collider operation. The requirement $| z_0 | < 60$~cm ensures that the cosmic-ray tracks  
 have similar trajectories as the particles selected for physics analysis.
 The data have been corrected for the known energy loss $\epsilon$. 
Also shown are the fits to Eq.~\ref{deltaPlusEquation} and the values and statistical uncertainties 
 of the fitted parameters (left) $a_2$ (in MeV), and $b_3$ (in GeV$^2$), and (right) $b_3$. 
 The error bars indicate the statistical uncertainties on the data points. The horizontal arrows indicate the range of
 $q/p_T$ of the leptons originating from $W^\pm \to \ell^\pm \nu$ and $Z \to \ell^+ \ell^-$ decays that are used in the $m_W$ 
 analysis~\cite{CDF2022}.}
\label{deltaPlusReduced}
\end{figure}

 With only $a_2$ and $b_3$ as free parameters, the fit to $\Delta^+_c$ as a function of $c_{\rm d}$ provides more information
 on these parameters, as shown in Fig.~\ref{deltaPlusReduced}. The $\chi^2$/dof is improved relative to Fig.~\ref{deltaPlus}, confirming 
 that $a_0$, $b_1$ and $\epsilon$ are redundant parameters per the AIC. 

The second-order bias in the $\Upsilon \to \mu \mu$ mass fit, with a typical $p_T \sim 5$~GeV, is  
\begin{itemize}
\item $a_0$ term: ($11 ~ {\rm PeV}^{-1} \cdot 5$~GeV)$^2 \sim$ (55~ppm)$^2 \sim 3$~ppb
\item $b_1$ term: (0.5 \permil)$^2 \sim 0.3$~ppm
\item $a_2$ term: $(\frac{3 ~ {\rm MeV}}{5 ~ {\rm GeV}})^2 \sim 0.4$~ppm
\item $b_3$ term: $[(0.036 ~ {\rm GeV}^2) (5~ {\rm GeV})^{-2}]^2 \sim (1.4~\permil)^2 \sim 2$~ppm
\end{itemize}
Note that the correlation coefficient between the fitted values of $a_2$ and $b_3$ in Fig.~\ref{deltaPlusReduced} (left) is $-97$\%, which
 means their combined bias is limited to 1.6~ppm. The uncertainties due to $a_0$, $b_1$ and $a_2$ are negligible compared to the total 
 quoted uncertainty on the momentum calibration of 25~ppm~\cite{CDF2022}, even without including their large anti-correlation, further 
 justifying their elimination from the fit of Fig.~\ref{deltaPlusReduced}.

Since the largest effect is due to $b_3$, the fit to $\Delta^+_c$ as a function of $c_{\rm d}$ 
 with only $b_3$ as the free parameter is shown in Fig.~\ref{deltaPlusReduced} (right). A small bias $b_3 = (0.023 \pm 0.008)$~GeV$^2$ 
 is discernible, corresponding to negligible bias of 1~ppm on the $\Upsilon \to \mu \mu$ mass fit.   
\subsubsection{$2^{\rm nd}$-order effects at the $m_{J/\psi}$ scale}
  \label{sec:mJ2}
A similar analysis for the typical $p_T \sim 3.3$~GeV in the $J/\psi \to \mu \mu$ data sample shows the following second-order mass bias:
\begin{itemize}
\item $a_0$ term: ($11 ~ {\rm PeV}^{-1} \cdot 3.3$~GeV)$^2 \sim$ (37~ppm)$^2 \sim 1$~ppb
\item $b_1$ term: (0.5 \permil)$^2 \sim 0.3$~ppm
\item $a_2$ term: $(\frac{3 ~ {\rm MeV}}{3.3 ~ {\rm GeV}})^2 \sim 0.8$~ppm
\item $b_3$ term: $[(0.023 ~ {\rm GeV}^2) (3.3~ {\rm GeV})^{-2}]^2 \sim (2.1~\permil)^2 \sim 5$~ppm
\end{itemize} 
Thus, the cosmic-ray data constrain  
  the possible bias in the $J/\psi \to \mu \mu$ data due to the $b_3$ term to 5~ppm which is much smaller than
  the systematic uncertainty of 25~ppm quoted for
 the momentum calibration uncertainty from these data~\cite{CDF2022}. 

 We conclude that the constraints on the ``$B$'' parameters of Eq.~\ref{eq:B} from cosmic-ray data result in negligible 
 uncertainty at all mass scales. 
\subsection{Additional bounds on $a_3$ and $b_3$ from $J/\psi$ and $\Upsilon$ data}
\label{sec:mass3}
\hspace*{0.06in}
The cosmic-ray data have a curvature range $0 < |c|~{\rm GeV} < 0.1$ which limits the precision on the $b_3$ parameter whose 
 effects are only visible at high curvature. The $J/\psi \to \mu \mu$ and $\Upsilon \to \mu \mu$ data extend the curvature range to 
 $0.2 < |c|~{\rm GeV} < 0.45$ and provide additional bounds on the $a_3$ and $b_3$ parameters. 
\subsubsection{Bound on $a_3$}
\label{sec:a3}
\hspace*{0.06in}
 The comparison of the mass fits to the $\Upsilon \to \mu \mu$ and $J/\psi \to \mu \mu$ data yields the following information. 
 The fractional bias on $m_{\Upsilon}$ ($m_{J/\psi}$) due to $a_3$ is $5^{-2}a_3$ ($3.3^{-2}a_3$). The difference between the momentum 
 calibrations extracted from these data is 20~ppm, consistent with the uncertainty on this difference~\cite{CDF2022}.  This implies
  that $a_3(3.3^{-2} - 5^{-2}) \lesssim 20$~ppm, which 
 bounds the uncertainty in extrapolating to $m_W$ to $a_3(5^{-2} - 40^{-2})$ to 15~ppm.   

 As the impact of $a_3$ grows with curvature, the range of $|c|$ spanned by the $J/\psi \to \mu \mu$ data yields additional information. 
 The momentum calibration in the bins of largest $|c| \sim 0.43$~GeV$^{-1}$ differs from the momentum calibration derived from the remainder of the 
 bins (whose typical $|c| \sim 0.3$~GeV$^{-1}$) by 20~ppm~\cite{CDF2022}. This implies that $a_3(0.43^{2} - 0.3^{2}) \lesssim 20$~ppm, which
 bounds the uncertainty in extrapolating to $m_W$ to $a_3(0.3^{2} - 40^{-2})$ to 19~ppm. Note that this small deviation observed in 
 the largest $|c|$ bins is the source of the largest systematic uncertainty quoted in the momentum calibration from the $J/\psi \to \mu \mu$ data in the CDF $m_W$ analysis~\cite{CDF2022}. 

 These two constraints together bound any effect of $a_3$ to 12~ppm. From the data presented in~\cite{CDF2022} (Fig.~2), one notes that 
 an attempt to extract $a_3$ from each of the above comparisons would yield two values of opposite sign. Thus the combined value of 
 $a_3$ would be much smaller than its estimated uncertainty of 12~ppm. 

 We conclude that, while the $J/\psi \to \mu \mu$ data and 
 the $\Upsilon \to \mu \mu$ data  together suggest a vanishing value of $a_3$, the rather tight bound on it is already included as an
 uncertainty on the momentum calibration in~\cite{CDF2022}.

\subsubsection{Bound on $b_3$}
\label{sec:b3}
\hspace*{0.06in}
 The comparison of the mass fits to the $\Upsilon \to \mu \mu$ and $J/\psi \to \mu \mu$ data can also yield information on $b_3$.
 The fractional bias on $m_{\Upsilon}$ ($m_{J/\psi}$) due to $b_3$ is $5^{-4}b_3^2$ ($3.3^{-4}b_3^2$). The consistency of these
 calibrations implies
  that $b_3^2(3.3^{-4} - 5^{-4}) \lesssim 20$~ppm, which 
 bounds the uncertainty in extrapolating to $m_W$ to $b_3^2(5^{-4} - 40^{-4})$ to 5~ppm.   

As the impact of $b_3$ grows rapidly with curvature, the range of $|c|$ spanned by the $J/\psi \to \mu \mu$ data yields additional 
 information on $b_3$. As mentioned above, the momentum calibration in the bins of largest $|c| \sim 0.43$~GeV$^{-1}$ differs from the 
 momentum calibration derived from the remainder of the
 bins (whose typical $|c| \sim 0.3$~GeV$^{-1}$) by 20~ppm~\cite{CDF2022}. This implies that $b_3^2(0.43^{4} - 0.3^{4}) \lesssim 20$~ppm, 
 which bounds the uncertainty in extrapolating to $m_W$ to $b_3^2(0.3^{4} - 40^{-4})$ to 6~ppm. 

As with the bounds on $a_3$, the $J/\psi$ and $\Upsilon$ data shown in~\cite{CDF2022} exert pulls on $b_3$ in opposite directions; their
 average yields a vanishing value of $b_3$, with a negligible uncertainty of 4~ppm. When combined with the constraint from the cosmic-ray 
 data of 5~ppm, the uncertainty due to $b_3$ reduces further to 3~ppm. 

\subsubsection{Summary of $a_3$ and $b_3$}
\hspace*{0.06in}
 The quoted uncertainty of 25~ppm on $m_W$ ascribed to the momentum calibration includes the potential uncertainty of 12~ppm from $a_3$, 
 and the uncertainty due to $b_3$ is negligible. 
\section{Application to the $m_W$ measurement}
\label{sec:wmass}
\hspace*{0.06in}
 In the $p \bar{p}$ collisions at the Tevatron, the beams of equal energy produce $W^+$ and $W^-$ bosons with identical momentum
 distributions except for a longitudinally antisymmetric component due to the $CP$-invariant initial state. The antisymmetric component 
 of $\Delta_{pe}$ is discussed and incorporated into the $m_W$ analysis~\cite{CDF2022}. Taking advantage of the ensuing symmetries in 
 the charged lepton distributions, as well as the longitudinal and azimuthal symmetry in the construction of the COT, we
 exploit the identity of the response function for both charges. 
\subsection{First-order bias in $m_W$}
\label{sec:wmass1}
\hspace*{0.06in}   
The $W$ boson mass is extracted from a fit to the $p_T$ distribution of the charged lepton,
 or to the distribution of transverse mass or $p_T^\nu$. The latter depend also on the 
 calibration of the hadronic recoil vector, which is calibrated using the tracker as reference. Thus we can consider $m_W \propto p_T \propto q/c$ and 
 the first-order fractional bias in the $W$ boson mass is 
$$\frac{- \delta m_W}{m_W} = q \frac{\delta c}{c^2} \frac{c}{q} = \frac{\delta c}{c} = \frac{a_0}{c} + (a_1 + b_1q) + (a_2 + b_2^\prime q)c 
 +  (a_3 + b_3q)c^2 $$
where the $\epsilon^2$ term is accounted for in the $m_W$ measurement with vanishing uncertainty and therefore dropped. 

Consider the half-difference of $\frac{\delta m_W}{m_W}$ between $W^-$ and $W^+$ bosons,
\begin{eqnarray}
  \Delta_W^\mp \equiv \frac{-1}{2}\Sigma_q q\frac{\delta m_W}{m_W} = \frac{1}{2}\Sigma_q q[\frac{a_0}{c} + (a_1 + b_1q) + (a_2 + b_2^\prime q)c +  (a_3 + b_3q)c^2]  \nonumber \\
   = \frac{1}{2}\Sigma_q [a_0p_T + b_1 + a_2p_T^{-1} + b_3p_T^{-2}] = a_0p_T + b_1 + a_2 p_T^{-1} + b_3 p_T^{-2} = B
\label{eq:deltaW}
\end{eqnarray}
\noindent 
since $a_1\Sigma_q q$, $b_2^\prime \Sigma_q c$ and $a_3 \Sigma_q qc^2$ vanish by charge symmetry. 
  This expression is identical to $\Delta_{pe}$ shown in Sec.~\ref{sec:eop}. As the  median $p_T$ and the Jacobian edge of the $p_T$ 
 distribution are similar within a few GeV, $\Delta_W^\mp$ is calibrated away by the 
 constraint from $\Delta_{pe}$, the positron-electron difference of $\langle E/p \rangle$.
 As expected, the charge-antisymmetric terms in $B$ (Eq.~\ref{eq:B}) provide the causal explanation for the $\Delta_W^\mp$ observable. 

We consider the average of $\frac{\delta m_W}{m_W}$ for $W^+$ and $W^-$ bosons, which reflects a first-order bias in $m_W$, 
$$\frac{1}{2}\Sigma_q \frac{\delta m_W}{m_W} = \frac{1}{2}\Sigma_q [\frac{a_0}{c} + (a_1 + b_1q) + (a_2 + b_2^\prime q)c + (a_3 + b_3q)c^2]
 = a_1 + b_2^\prime p_T^{-1} + a_3 p_T^{-2} = A$$
\noindent
because $a_0 \langle c^{-1} \rangle$, $b_1 \Sigma_q q$, $a_2 \langle c \rangle$ and $b_3 \langle qc^2 \rangle$ vanish by charge symmetry. 
 As expected, the charge-symmetric terms in $A$ (Eq.~\ref{eq:A}) induce an $m_W$ bias.

  As mentioned in Sec.~\ref{sec:mass}, $b_2^\prime$ is constrained to within 34 keV, hence its contribution
 to a first-order bias in $m_W$ is negligible ($b_2^\prime \langle p_T^{-1}\rangle \sim \frac{34~{\rm keV}}{35~{\rm GeV}} \sim 1$~ppm). 
 The momentum scale parameter $a_1$ has been calibrated to  25~ppm~\cite{CDF2022}, 
 which includes a potential uncertainty due to $a_3$ of 12~ppm. We conclude that the analytic curvature response function, 
 expanded up to $c^3$ terms, is sufficiently well-constrained to prevent significant first-order bias in $m_W$.  
\subsection{Second-order bias in $m_W$}
\label{sec:wmass2}
\hspace*{0.06in}
Extending  the derivative to second-order, 
$$\frac{\delta m_W}{m_W} = q \frac{(\delta c)^2}{c^3} \frac{c}{q} = (\frac{\delta c}{c})^2$$
\noindent
which after calibration (i.e. dropping the $a_1$, $b_2^\prime$ and $\epsilon^2$ terms) simplifies to
$$\frac{\delta m_W}{m_W} = (a_0c^{-1} + b_1q + a_2c + a_3c^2 + b_3qc^2)^2 
 = (a_0p_T + b_1 + a_2p_T^{-1} + b_3p_T^{-2} + a_3qp_T^{-2})^2$$
$$ = (\Delta_W^\mp + a_3qp_T^{-2})^2 = (\Delta_W^\mp)^2 + (a_3 p_T^{-2})^2 + 2q(\Delta_W^\mp)(a_3p_T^{-2})$$
\noindent
 The quantity $\Delta_W^\mp = (m_{W^-} - m_{W^+})/m_W$ is measured to be consistent with zero within a statistical precision of 0.2\permil~\cite{CDF2022}, therefore 
 the first term
 is $0.04$~ppm based on this measurement. If $\Delta_{pe}$ is used instead of $\Delta_W^\mp$, the constraint has a statistical precision of (43~ppm)$^2 = 0.002$~ppm which 
 is tighter by another order of magnitude. As mentioned in Sec.~\ref{sec:mWZ2}, the constraints from $\Delta_W^\mp$ and $\Delta_{pe}$ apply
  mostly to the low-curvature parameters $a_0$ and $b_1$; this is reflected in Table~\ref{tab:parameterSummary}. 

 The term $a_3 p_T^{-2}$ is constrained to 12~ppm from the $J/\psi$ and $\Upsilon$ data, hence the square of this term is vanishing. 

 The third (product) term is charge-dependent, of $\cal O$$(2 \times 0.2\permil \times 34~{\rm ppm}) \sim 0.01$~ppm or smaller, 
 which is vanishing. 

 We conclude that the second-order bias on $m_W$ can be estimated from the constraints on the parameters in the COT response function, from
 the measurement of $\Delta_W^\mp$, and from the calibration based on $\Delta_{pe}$. All estimates are vanishingly small.   
\subsection{Summary of analytic curvature response function}
\label{sec:wmassTot}
\hspace*{0.06in}
The parameters of the analytic curvature response function (Eq.~\ref{eq:analytic}) propagate to the momentum calibration uncertainty
 on $m_W$ at first- or second-order. Based on cosmic-ray data and the published $J/\psi$ and $\Upsilon$ data~\cite{CDF2022}, all 
 uncertainties are either already included in the  $m_W$ analysis~\cite{CDF2022} or found to be negligible.

 As mentioned earlier,  the CDF procedure to calibrate track $p_T$ is based on mass measurements of the $J/\psi$ and $\Upsilon$
 mesons in the dimuon channel~\cite{CDF2022,CDF2firstPRD,CDF2014}. These data samples are binned in the mean $|c|$ of the two muons and mass fits
 are performed in each bin separately. An inclusive mass fit is also performed for the $\Upsilon$ sample. The mass fits are based on likelihood maximization
 using simulated templates for the signal line-shapes. The background shapes are modeled and their normalizations are constrained using 
 the sidebands of each mass peak in the data. The simulation models meson/boson production and decay and the propagation of muons (and electrons) 
 through the detector, including energy loss and resolution effects. The mass bias $\delta m$ is defined as the difference of the measured meson mass 
 from the world-average reference (particle data group) value.
 
 We have shown that the fractional mass bias $\delta m/m$ as a function of $|c|$ is parameterized in general by the $A$ terms (Eq.~\ref{eq:A}) 
 at first order, and by $B^2$ (Eq.~\ref{eq:B})  at second order. We have also shown that $B^2$ is strongly constrained to be zero by both the cosmic-ray
 analysis and by $\Delta_{pe}$, the positron-electron difference of $E/p$ (Eq.~\ref{eq:deltaPE}). Since they contribute negligibly, the $B^2$ terms can be dropped in
 subsequent analysis and only the $A$ terms need to be extracted. Furthermore, since the $a_3$ term in $A$ is
 redundant within the quoted uncertainties on $a_1$ and $b_2^\prime$, it is appropriate to let $a_3 = 0$ and extract $a_1$ and $b_2^\prime$ from a linear fit to 
 $\delta m/m$ versus $|c|$, as shown in Fig.~2A of~\cite{CDF2022}, Fig.~13 of~\cite{CDF2014} and Fig.~22 of~\cite{CDF2firstPRD}. 
 In practice, the energy loss $\epsilon$ is tuned in the simulation by a few \% of its {\it ab initio} value such that the linear fit returns $b_2^\prime = 0$.
 Therefore these figures in the respective publications display the value of $a_1$ corresponding to the as-built detector and track reconstruction
 software, i.e., prior to final calibration of track $p_T$  at the analysis level.

 We note that the four points from $J/\psi \to \mu \mu$ data at small $|c|$ in Fig.~2A of~\cite{CDF2022}, if combined, would deviate
 from the model by $2 \sigma$. The large number of points that are consistent with the model, together with the stringent constraints on the model from multiple independent
 control samples of data, preclude non-linear models. In order for these points to be indicative of a systematic effect,
 the model must include the $B^2$ term of the form $(a_0p_T)^2 = (a_0/c)^2$. We have shown that this term  has been eliminated by the high-quality alignment, both internal and
  external, of the COT, and by the $\Delta_{pe}$ constraint. Furthermore, this term
 has negligible impact in the range of $|c|$ where the $J/\psi$ and $\Upsilon$ data constrain $a_1$. Hence the extraction of $a_1$ from the $J/\psi$ and $\Upsilon$ data using
  the linear fit to $\delta m/m$ versus $|c|$ is robust. There is no justification for a non-linear fit, given the preponderance of evidence against it, merely to accommodate a fluctuation. 
 
 These procedures lead to the inference that $\delta m/m = a_1$ for all resonances. CDF developed  these procedures to constrain $a_1$ from the $J/\psi$ and $\Upsilon$ data and
 apply it to calibrate the $p_T$ of tracks from $W$ and $Z$ boson decays at the analysis level. The uncertainties due to $b_2^\prime$ and $a_3$ have been included
 in the quoted uncertainty on $a_1$, and we have shown that all other terms in the curvature response function (Eq.~\ref{eq:analytic}) are negligible. Thus, the most
 general analytic response function is completely pinned down. 
\section{Singular curvature response functions}
\label{sec:singular}
\hspace*{0.06in}
Equation~\ref{eq:analytic} shows the analytic response function for curvature as a Maclaurin expansion, appropriate 
 for the $W^\pm$, $Z$, $\Upsilon$ and $J/\psi$ samples which have symmetric $c$ distributions about $c = 0$ at the 
 Tevatron\footnote{The leptons from $W$ boson decays do not have symmetric $c$ distributions at the LHC.}. 
 Since tracking detectors measure 
 curvature, the coefficients of the response function relate to the geometry and the physics principles on which the hit measurements are 
 based. 
\subsection{Terms with negative exponents}
\label{sec:Laurent}
\hspace*{0.06in}
One may consider extending the response function from an analytic Maclaurin series to a non-analytic Laurent series by adding terms 
 of the form $c^{-|n|}$ for integer values of $n$. Such terms must be forbidden on physical grounds; as $c \to 0$ they 
 imply $\delta c \to \infty$ which means that the particle's straight-line trajectory becomes indeterminate. Equivalently, such 
 terms are inconsistent with the simple fact that if the axial magnetic field were ramped down to zero magnitude, $B_z \to 0^\pm$, 
 all trajectories tend to straight lines which can easily be reconstructed by the drift chamber with no loss of directional information. 

\begin{figure}[h] \begin{center} \includegraphics[width=6.4in]{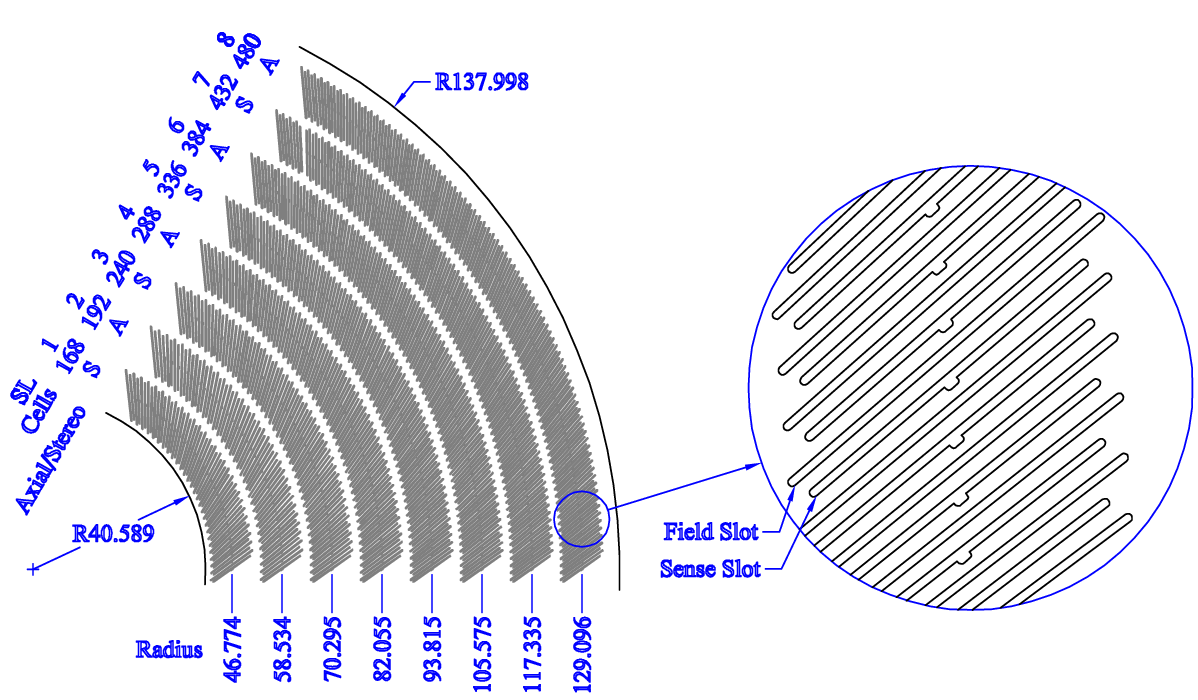} \end{center}
\caption{End view of a section of the CDF COT end plate. The sense wires are organized into  eight concentric ``superlayers''. Each
 superlayer is partitioned  
 azimuthally into cells, and each cell contains 12 sense wires separated from adjacent cells by field sheets. Precision-machined slots in
 the end plates hold each cell's sense wires and field sheets under tension. The radius at the center of each superlayer is shown in cm.
 Figure reproduced with permission from Fig.~2 of~\cite{cotNim}.} \label{fig:COTquad}
\end{figure}

Figure~\ref{fig:COTquad} shows the physical construction of the COT~\cite{cotNim}. It is a single, cylindrical gaseous volume in which 
 sense wires
 and field sheets are embedded. The only way for a particle trajectory to be indeterminate is if the particle were to traverse
 an uninstrumented region of the tracking volume. As $c \to 0$ the particle trajectories approach a straight line in the radial
 direction. All sense-wire and field-sheet planes are tilted (at $35^{\rm o}$) with respect to the radial direction in an azimuthally symmetric manner~\cite{cotNim}. 
 With this construction, there is no dead space that the particle can traverse; the entire volume is active. There are no azimuthal 
 boundaries through which a particle may pass without ionizing gas adjacent to every sense wire. Hence, every radial 
 straight line is detectable as such. 
\subsubsection{COT performance studies with cosmic rays}
\label{sec:COTperf}
\hspace*{0.06in}
As described in Sec.~\ref{sec:cosmics}, the cosmic-ray sample is collected in situ with collider data using the high-$p_T$ inclusive muon 
trigger that also acquires the $W \to \mu \nu$ and $Z \to \mu \mu$ data. The sample can therefore be used to study the COT performance under the 
same operating conditions as the physics signals.

\begin{figure}[h] \begin{center} 
\includegraphics[width=3.5in]{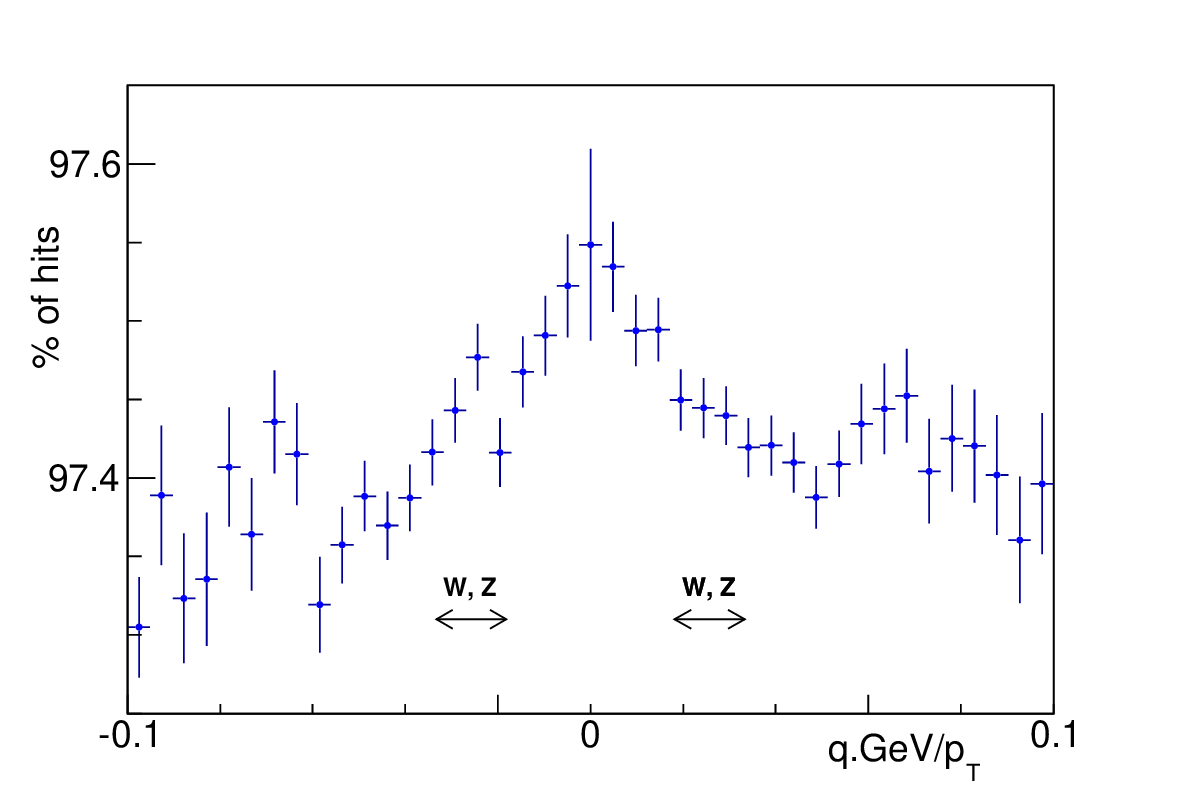} 
\includegraphics[width=3.5in]{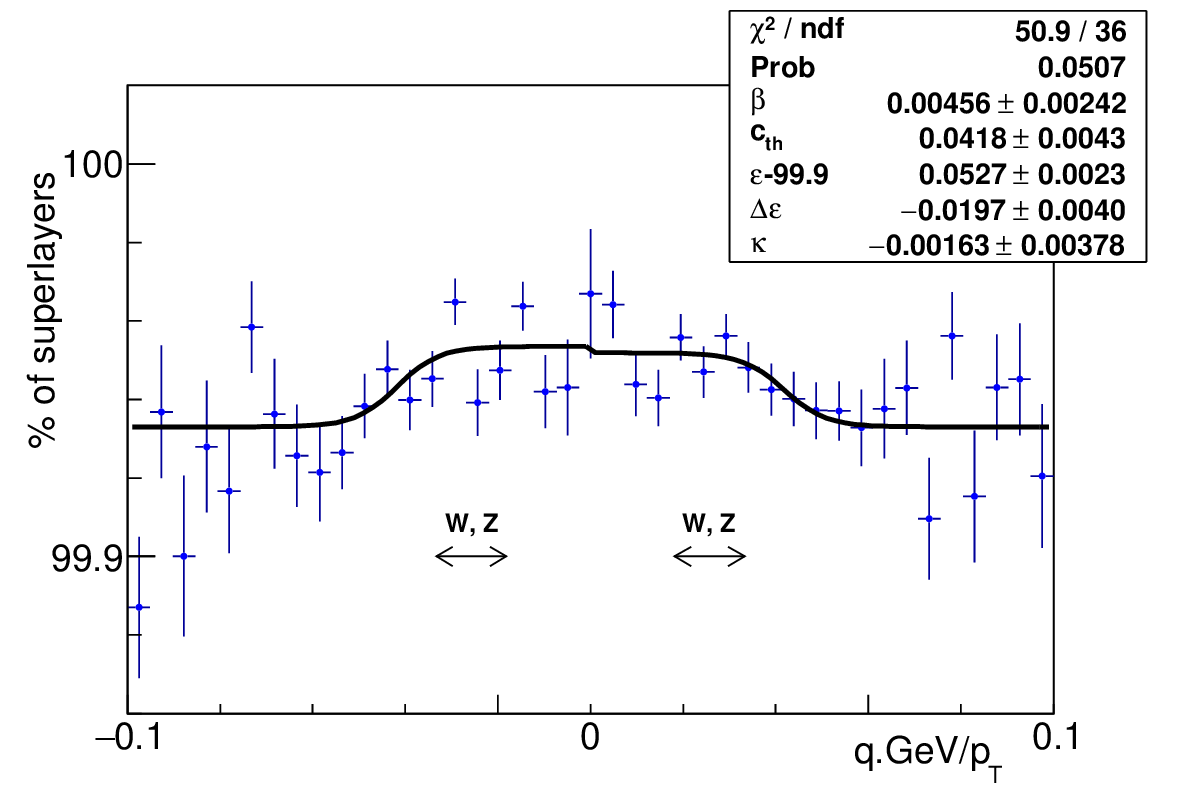} 
\end{center}
\caption{(Left) The fraction of the maximum possible number of COT hits contributing to the cosmic-ray tracks. (Right) The fraction of 
 the  maximum possible number of superlayers contributing to the cosmic-ray tracks, when a superlayer is required to contribute at least 
 5 (out of 12) hits. The fit is intended to guide the eye. The superlayer efficiency is constant at 99.95\% at low curvature and lower by
 0.02\% for $|c_{\rm th}^{-1}| = p_T \lesssim 25$~GeV when a different trigger is used. 
 The fit function is $y = \varepsilon + \Delta \varepsilon/(1 + e^{-\xi}) + \frac{q}{2}\kappa/(1 + e^{\xi})$ where $\xi = (|x|-|c_{\rm th}|)/\beta$.
 The parameter $\kappa = (0.002 \pm 0.004)$\% describes the difference in superlayer efficiency as $c \to 0^+$ versus $c \to 0^-$.  
} \label{fig:COThits}
\end{figure}

The fraction of the maximum possible number of 
 COT hits contributing to cosmic-ray tracks is shown in Fig.~\ref{fig:COThits}. As the track tends to a straight 
 line, the fraction of hits associated with the track increases slightly, by 1\permil, because pattern recognition becomes easier as the 
 curvature reduces. Importantly, there is no indication of a loss of performance or efficiency in the 
 $c \to 0^\pm$ limit. 

 The typical alignment accuracy of the wires is 1~$\mu$m~\cite{cosmicAlignment,CDF2022}. Even if we make the extremely conservative assumption that the change in 
 efficiency is due to one particular wire which induces a curvature bias of $\cal O$(1~$\mu$m/$l^2$), where $l \sim 1$~m is the length
 of the track, this bias would amount to 0.2~ppm at $m_{W,Z}$ (see Appendix~\ref{appendixWireBias} for the calculation). Furthermore, 
 as a (signed) curvature bias, it cancels at first order in sign-averaged mass measurements and would be vanishing at second order. In
 order to induce a first-order bias, there would need to be a difference in this inefficiency variation between positive and negative 
 particles. As seen in Fig.~\ref{fig:COThits}, the difference between positive and negative tracks is much smaller than 1\permil. We 
 conclude that 
 the high degree of stability of the hit efficiency (pattern recognition) with respect to curvature, together with the accurate alignment, 
 supports the ansatz that the COT has no discernible discontinuity in the $c \to 0^\pm$ limit.   

 Another performance metric is the fraction of superlayers that contribute to the track (see Fig.~\ref{fig:COThits}), where a 
 superlayer is counted if it contributes at least 5 out of the maximum 12 hits. This criterion is the one used in the $m_W$ analysis
 for candidate track selection~\cite{CDF2022}. The superlayer fraction is $\varepsilon = 999.5$\permil, independent of curvature as 
 $c \to 0^\pm$ and reduces by $\Delta \varepsilon = 0.2$\permil\ for $p_T$ values below the selection criterion used for $W$ and $Z$ boson
  samples~\cite{CDF2022}. This metric is the most 
 relevant for the curvature response function and indicates that the COT performance is independent of curvature in the $c \to 0$ limit. 

 In Fig.~\ref{fig:COThits} (right), the parameter $\kappa$ is designed to capture any charge-dependent superlayer efficiency as $c \to 0^\pm$.
 We find that $\kappa$ is zero within a statistical precision of 0.04\permil, strongly disfavoring any discontinuous behavior.
 
 In Appendix~\ref{appendixWireBias} we also estimate the possible mass bias induced by the superlayer inefficiency, and arrive at a 
 more realistic bound of 4~ppb, based on a maximum superlayer efficiency variation of 0.1\permil\ in the $c \to 0^\pm$ limits. 

\begin{figure}[h] \begin{center} 
\includegraphics[width=3.5in]{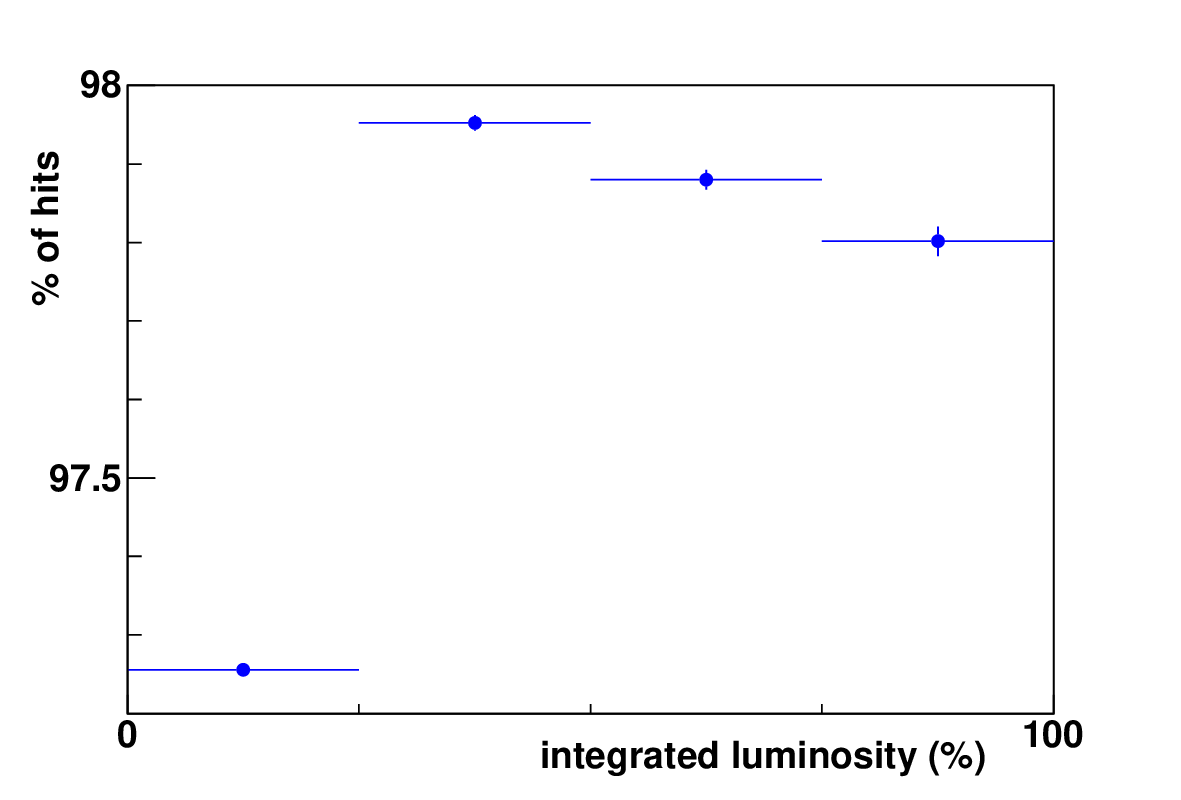} 
\includegraphics[width=3.5in]{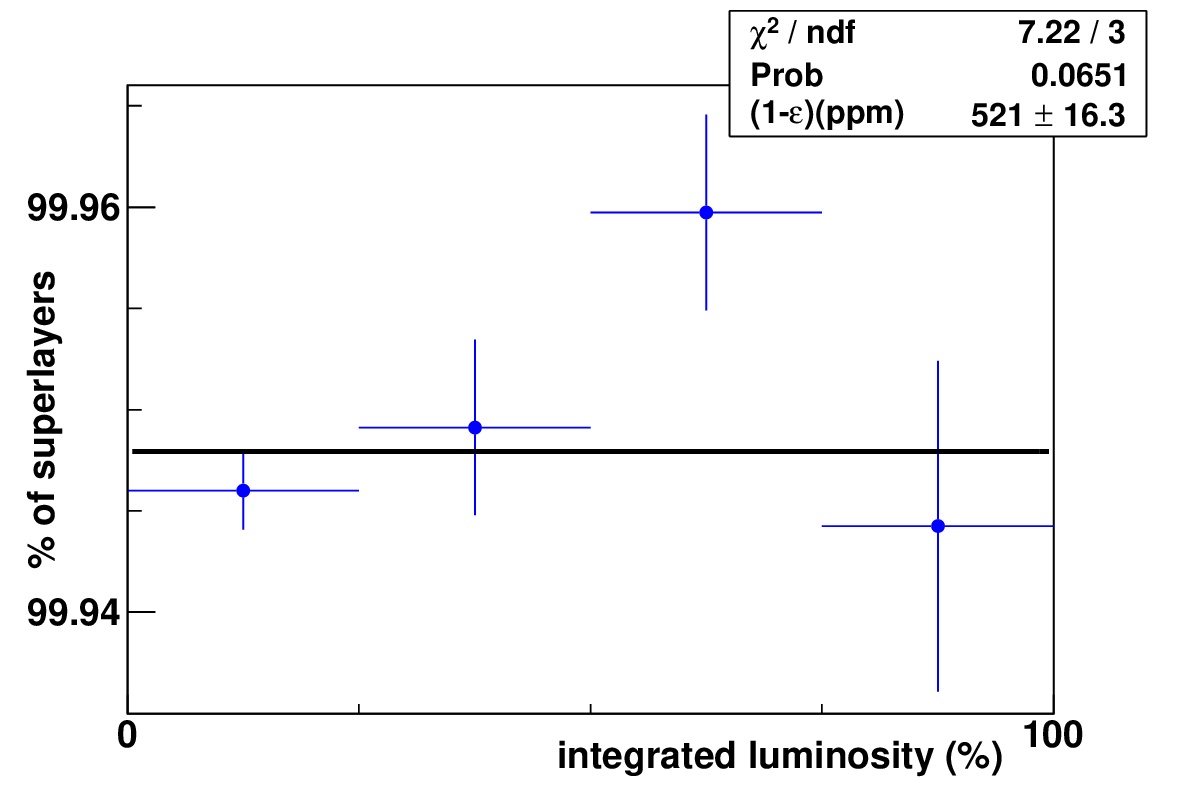} 
\end{center}
\caption{(Left) The fraction of the maximum possible number of COT hits contributing to the cosmic-ray tracks. (Right) The fraction of 
 the  maximum possible number of superlayers contributing to the cosmic-ray tracks, when a superlayer is required to contribute at least 
 5 (out of 12) hits. The fractions are shown in four sequential time periods, with each period delivering a quarter of the total 
 integrated luminosity recorded by CDF II and used for the $m_W$ measurement~\cite{CDF2022}. The superlayer inefficiency is consistent
 with a constant value of $(520 \pm 30)$~ppm over the entire 10-year operation of the COT. For these data, the requirement 
 $p_T > 20$~GeV is imposed in order to be applicable to the $W$ and $Z$ boson data. The complementary sample with $10 < p_T < 20$~GeV 
 shows similarly stable efficiencies at slightly lower values, as seen in Fig.~\ref{fig:COThits}.   
} \label{fig:COThitsLumiBlock}
\end{figure}

The stability of the COT performance with respect to time is demonstrated in Fig.~\ref{fig:COThitsLumiBlock}. In order to interpret 
  time stability  in the context of collider data, hit and superlayer efficiencies are presented in sequential 
 blocks of integrated luminosity. Fig.~\ref{fig:COThitsLumiBlock} shows that the hit efficiency actually increased after 2007, when
 3/4 of the data were collected, and remained remarkably stable post-2007, with a worst-case drop of 1.5\permil. The superlayer efficiency is even more stable due to 
 the inherent redundancy in the requirement of 5-of-12 hits per superlayer. This 
 inefficiency is consistent with a constant value of $520 \pm 30$~ppm over the 10-year operation of the COT, proving that there is
 little evidence for degradation or radiation damage and no visible impact on tracking efficiency.

 Stability with respect to instantaneous luminosity, ie. occupancy is provided by the COT track reconstruction procedure. After the first
 pass of pattern recognition, hit association and track-fitting, a second-pass refit of COT tracks is performed. Prior to this refit, 
 hits whose distance from the track exceeds a threshold of about $4\times$ the hit resolution are removed from the track. Next,
 unused hits whose distance from the track is less that about $3\times$ the hit resolution are added to the list of associated hits.
 The updated hit list is refitted to obtain the track parameters used for analysis. This ``drop-add'' procedure removes spurious hits
 and rescues hits that were missed in the first-pass reconstruction due to high occupancy. Together with the high level of redundancy in
 the COT tracker (up to 96 hits per track), the procedure makes COT tracks robust,  stable and insensitive to occupancy.  
\subsubsection{Summary of performance studies and negative-exponent terms}
\label{sec:summaryLaurent}
\hspace*{0.06in}
By design, the COT geometry is continuous, without boundaries, and is sampled uniformly by tracks in the $c \to 0^\pm$ limits. 
This reasoning eliminates terms of the form $c^{-|n|}$ and by extension,  $c^{-|r|}$ for real values of $r$. Studies of hit efficiency 
and superlayer efficiency as a function of curvature support this reasoning at a more stringent level (by four orders of magnitude)
 than the uncertainty of 25~ppm 
 on the momentum calibration quoted in~\cite{CDF2022}. 
\subsection{Terms with fractional exponents}
\label{sec:Puiseux}
\hspace*{0.06in}
The Maclaurin series may be extended to a Puiseux series by including terms with fractional (positive) exponents.  
We consider terms with fractional positive powers of $c$ in the response function. Since $c$ is a signed quantity we must consider two types
 of terms, $|c|^r$ and $q |c|^r$ where $0 < r < 1$. In principle, such factors may multiply the entire analytic function of 
 Eq.~\ref{eq:analytic}. However, the $a_1$ and $\epsilon$ terms have obvious physical interpretations and may not be altered. 

The  degrees of freedom incorporated in Eq.~\ref{eq:analytic} already span the phase space of the calibration data. By the AIC we can 
 use the minimal model that adequately describes the data, and propagate the uncertainties in the model parameters. Therefore the 
 additional terms of fractional powers are only needed to explore the $c \to 0$ limit.  

 We consider the terms $a_r |c|^r$ and $b_r q |c|^r$ individually, since they are even and odd in $q$, respectively. 
\subsubsection{Charge-independent term with fractional exponent}
\label{sec:Ar}
\begin{figure}[h]
\begin{center}
\includegraphics[width=3.5in]{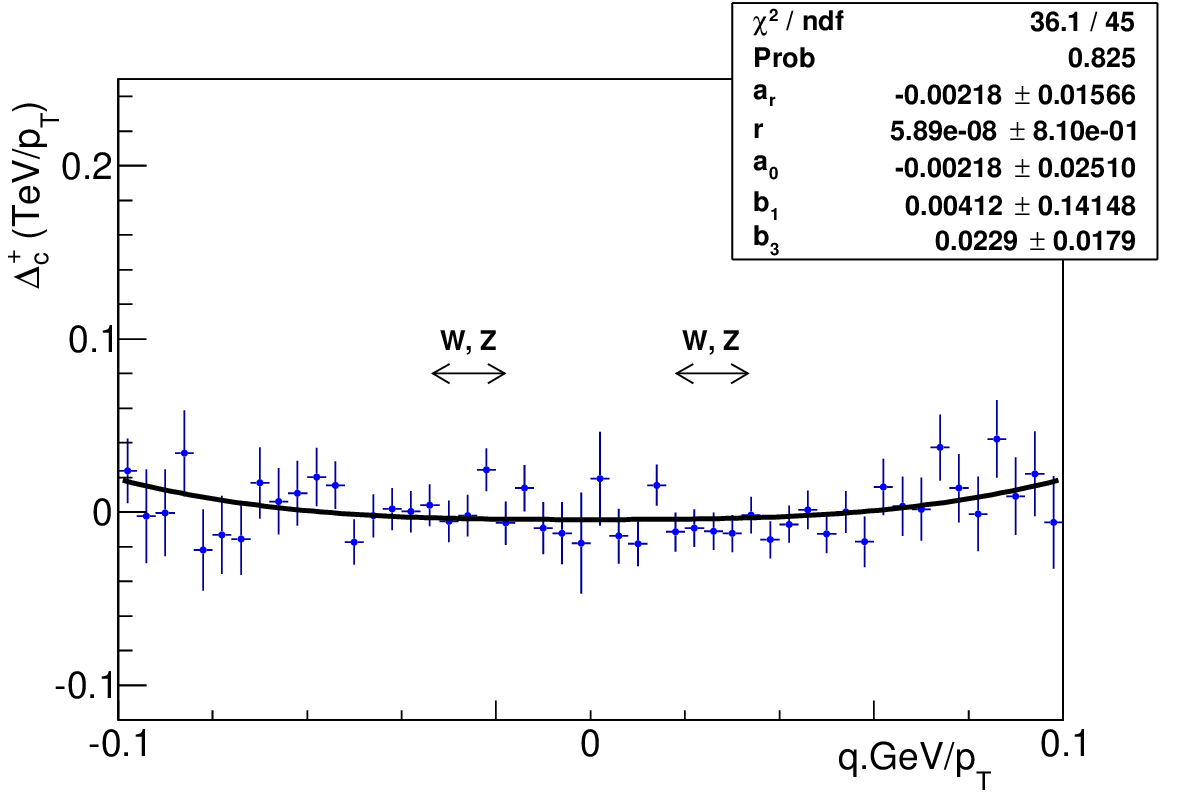}
\includegraphics[width=3.5in]{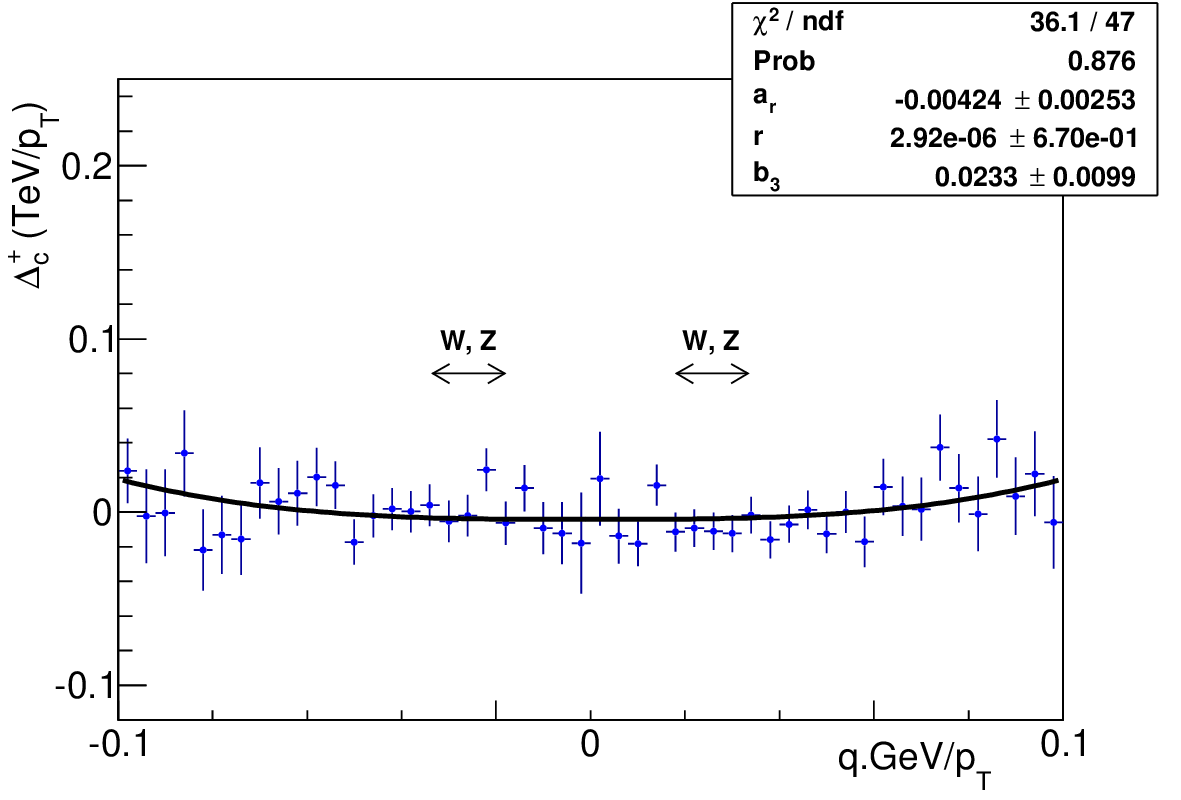}
\end{center}
\caption{(Left) The measurement of $\Delta^+_c$ as a function of $c_{\rm d}$, the measured curvature of the COT dicosmic helix, in 
 cosmic-ray data collected in situ during collider operation. The requirement $| z_0 | < 60$~cm ensures that the cosmic-ray tracks   
 have similar trajectories as the particles selected for physics analysis.  
 The data have been corrected for the known energy loss $\epsilon$, same as in Fig.~\ref{deltaPlusReduced}.  
Also shown is the fit to Eq.~\ref{deltaPlusEquation} including the $a_r |c|^r$ term ($0 < r < 1$), 
 with the values and statistical uncertainties   
 of the fitted parameters $a_r$, $r$, $a_0$ (in TeV$^{-1}$), $b_1$ (in \permil) and $b_3$ (in GeV$^2$). 
 (Right) The same data and fit with $a_r$, $r$ and $b_3$ as the fitted parameters.    
 The error bars indicate the statistical uncertainties on the data points. The horizontal arrows indicate the range of  
 $q/p_T$ of the leptons originating from $W \to \ell^\pm \nu$ and $Z \to \ell^+ \ell^-$ decays that are used in the $m_W$ 
 analysis~\cite{CDF2022}.}
\label{deltaPlusAr}
\end{figure}

\hspace*{0.06in}
In Fig.~\ref{deltaPlusAr} (left) we extend the fit of Fig.~\ref{deltaPlusReduced} to include the term $a_r |c|^r$. The fit finds 
 $r=0$ which means that this term is redundant with $a_0$; a conclusion confirmed by their mutual correlation coefficient of $-97$\%. 
 Also, the $b_1$ and $b_3$
 terms are anticorrelated with a correlation coefficient of $-88$\%. The $a_0$ and $b_1$ terms are dropped to increase the incisiveness
 of the fit; the $\chi^2$/dof decreases as shown in Fig.~\ref{deltaPlusAr} (right). This fit returns $r=0^{+0.06}_{-0}$ which implies 
 that the data do not differentiate this term from the $a_0$ term. 

\begin{figure}
\begin{center}
\includegraphics[width=3.5in]{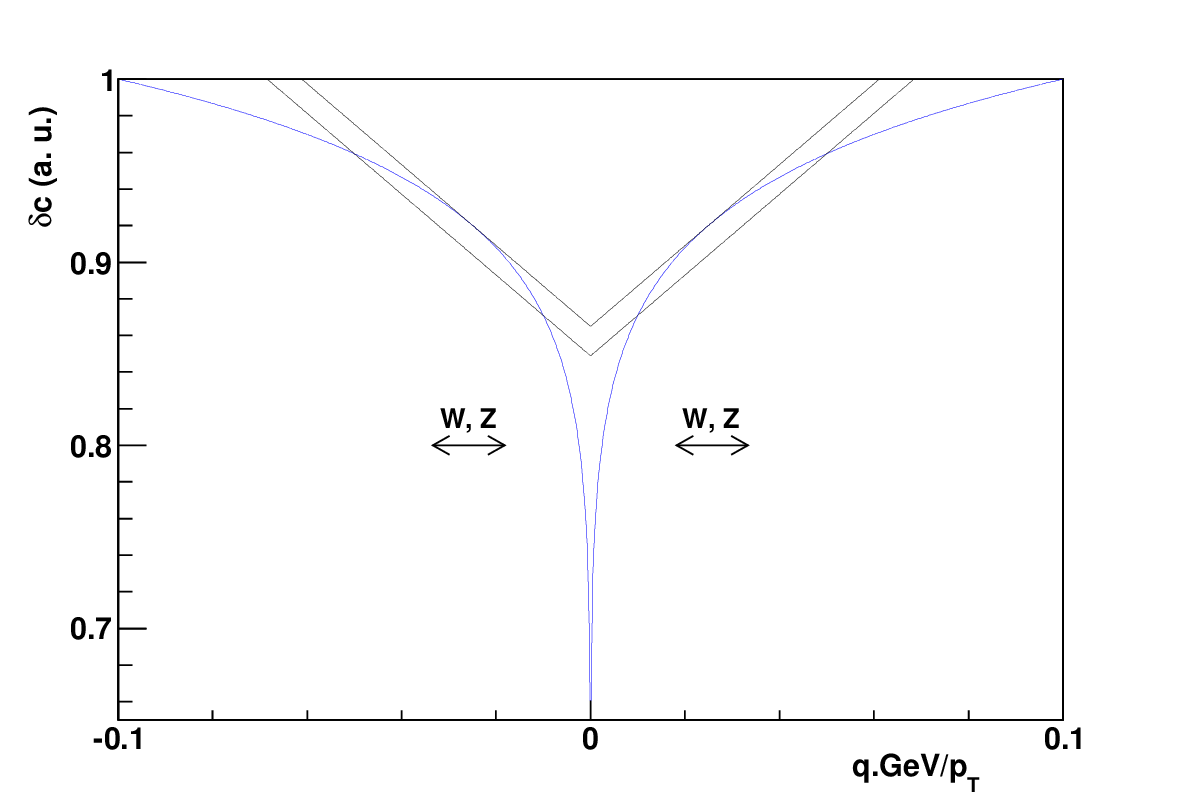}
\includegraphics[width=3.5in]{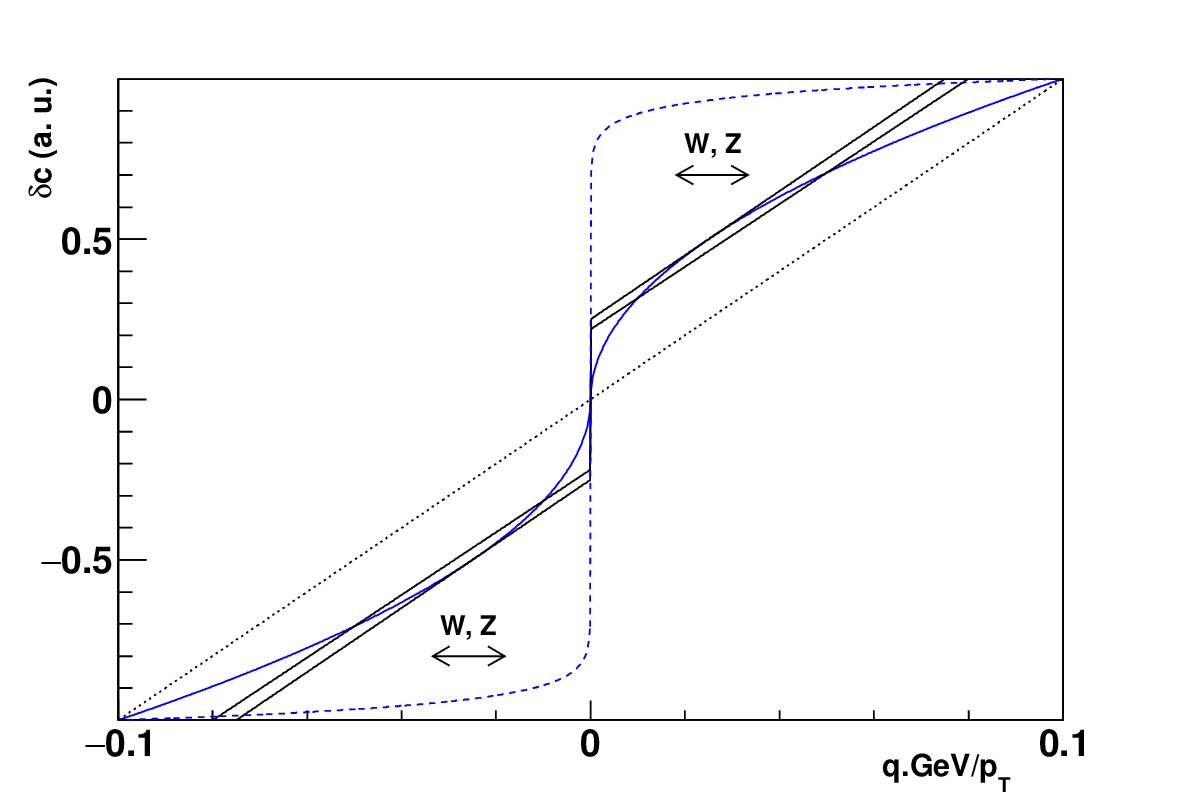}
\end{center}
\caption{(Left) An illustration of the function $|c/0.1|^{0.06}$ (solid blue curve).
  (Right) An illustration of the function $q|c/0.1|^r$ for $r=0.5$ (solid blue curve), $r=0.05$ (dashed blue curve), and $r=1$ 
 (dotted black line). In each illustration, two straight-line approximations are shown for the solid blue curve; a tangent at 
 $p_T = 40$~GeV and a line that intersects the curve at $p_T = 10$ and 100~GeV. The linear approximations (solid black lines) indicate that a fractional
 exponent term can be mimicked by an intercept, a slope and higher-order terms in the relevant range of curvature. }
\label{ArTerm}
\end{figure}

In Fig.~\ref{ArTerm} (left) we illustrate the function $|c/0.1|^{0.06}$ where the exponent is chosen to reflect the range compatible
 with the fit of Fig.~\ref{deltaPlusAr} (right). This term is highly non-differentiable at $c=0$; the slope changes from $-\infty$ at 
 $c = 0^-$ to $\infty$ at $c = 0^+$. It implies that zero curvature is distinguishable from vanishing curvature, a conclusion  
 inconsistent with the fact that all drift distances in the COT change continuously with $c$ for all relevant values of $c$. There
 is no physical model that generates such a fractional-exponent term. 

Also shown in Fig.~\ref{ArTerm} (left) are two linear functions based on the $a_0$ and $b_1$ terms from Eq.~\ref{eq:analytic}. 
 These linear functions are intended to guide the eye. One of these lines intersects $|c/0.1|^{0.06}$ at $|c| = 0.05$~GeV$^{-1}$ and 
 $|c| = 0.01$~GeV$^{-1}$, 
 i.e. $p_T = 20$~GeV and 100~GeV which flank the range of the charged leptons from $W$ and $Z$ boson decays. The second line 
 is tangent to $|c/0.1|^{0.06}$ at $p_T = 40$~GeV, typical for these leptons. The similarity of these linear functions to $|c/0.1|^{0.06}$
 indicates that this function can be adequately approximated by Eq.~\ref{eq:analytic} in the relevant $p_T$ range; furthermore, the 
 approximation improves as the exponent reduces. 

We conclude that a charge-independent term with  fractional exponent is excluded by the cosmic-ray data. It is also inconsistent
 with the known physical properties of the COT. In any case, the $a_0$, $b_1$, $a_2$ and $b_3$ terms provide a physically
 justifiable model that can adequately mimic any such effect in the relevant range of curvature. 
\subsubsection{Charge-dependent term with fractional exponent}
\label{sec:Br}
\hspace*{0.06in}
 We consider the extremes of the possible range  $0<r<1$ of the exponent in the term $b_r q |c|^r$. As $r \to 1$ the term 
 $b_r q |c|^r \to a_1 c$ and hence becomes redundant with the momentum scale. As $r \to 0$ the derivative of the response function near $c=0$ diverges more 
 rapidly as $|c|^{r-1}$. This is illustrated 
 in Fig.~\ref{ArTerm} (right). We also have $\lim_{r \to 0} b_r q |c|^r = b_0q$ so that the response function develops a discontinuous 
 step function at $c=0$. The implication is that a nearly straight track (with indeterminate $q$) is reconstructed with 
 a false curvature whose sign depends on $q$. A physical explanation for such a response function of the COT is not plausible on the 
 basis of its geometry and principles of operation, since
 the response cannot be insensitive to $q$ and depend strongly on $q$ at the same time. 
\subsubsection{Effect of singular term on observables}
\label{sec:singularImpact}
\hspace*{0.06in}
If the term $b_0q$ is added to the response function of Eq.~\ref{eq:analytic}, it cancels in the sum over the two legs of the cosmic ray
and does not show up in $\Delta^+_c$ (Eq.~\ref{deltaPlusEquation}).
It is  shown in Appendix~\ref{appendixBminus} that $\Delta^-_c$ cannot constrain this term.
 
The effect of the $b_0$ term on $\Delta_{pe}$, the positron-electron difference of $\langle E/p \rangle$ (Eq.~\ref{eq:deltaPE}), and
on $\Delta_W^\mp$, the mass difference between $W^-$ and $W^+$ bosons (Eq.~\ref{eq:deltaW}), is vanishing, 
  $$\frac{1}{2}\Sigma_q q \frac{\delta c}{c} = \frac{1}{2}\Sigma_q q \frac{b_0q}{c} = \frac{b_0}{2}\Sigma_q c^{-1} = 
 \frac{b_0}{2}\Sigma_q qp_T = 0$$ 
by charge symmetry. 

The effect of the $b_0$  term on the reconstructed invariant mass is 
$$\frac{\delta m}{ m } |_{\rm 1st} = \frac{-1}{2}\Sigma_q\frac{\delta c}{c} = \frac{-1}{2}\Sigma_q\frac{b_0q}{c} 
 = \frac{-b_0}{2}\Sigma_q\frac{q}{c} = \frac{-b_0}{2}\Sigma_q p_T = -b_0 \langle p_T \rangle$$
Thus, any COT measurement bias that is mimicked by the $b_0 q$ term will result in biased $m_W$  and $m_Z$ measurements   
 because it will not be calibrated by the $J/\psi$ and $\Upsilon$ data at lower $p_T$. 
\subsubsection{Study of discontinuous COT response}
\label{sec:b0qStudy}
\hspace*{0.06in}
As mentioned above, the $b_0 q$ term in the response function requires an unphysical discontinuity in the response as $c \to 0^\pm$. 
 Since  the entire gaseous volume is instrumented, a discontinuity would have to be generated by some feature of the electron drift. The 
 drift cells are build to be almost mirror-symmetric (the plane of sense wires is half-way between the parallel planes of field 
 sheets~\cite{cotNim}), so the fraction of hits that involve leftward and rightward drifts are roughly equal. A small asymmetry between
 these fractions can be generated by the $35^{\rm o}$ supercell\footnote{A supercell consists of 12 drift cells, with each cell
 containing one sense wire.} tilt which compensates for the Lorentz angle in the magnetic field~\cite{cotNim}, and by the 
 deflection of the wires due to the electrostatic forces from the field sheets~\cite{cotNim,cosmicAlignment}. 

 Related to the left-right drift asymmetry is the average displacement of the hits from the sense wires. For exactly left-right symmetric
 drift, the average hit distance from the sense wire would be equal in both directions, thereby canceling in the average displacement. 
 
\begin{figure}[t]
\begin{center}
\includegraphics[width=3.5in]{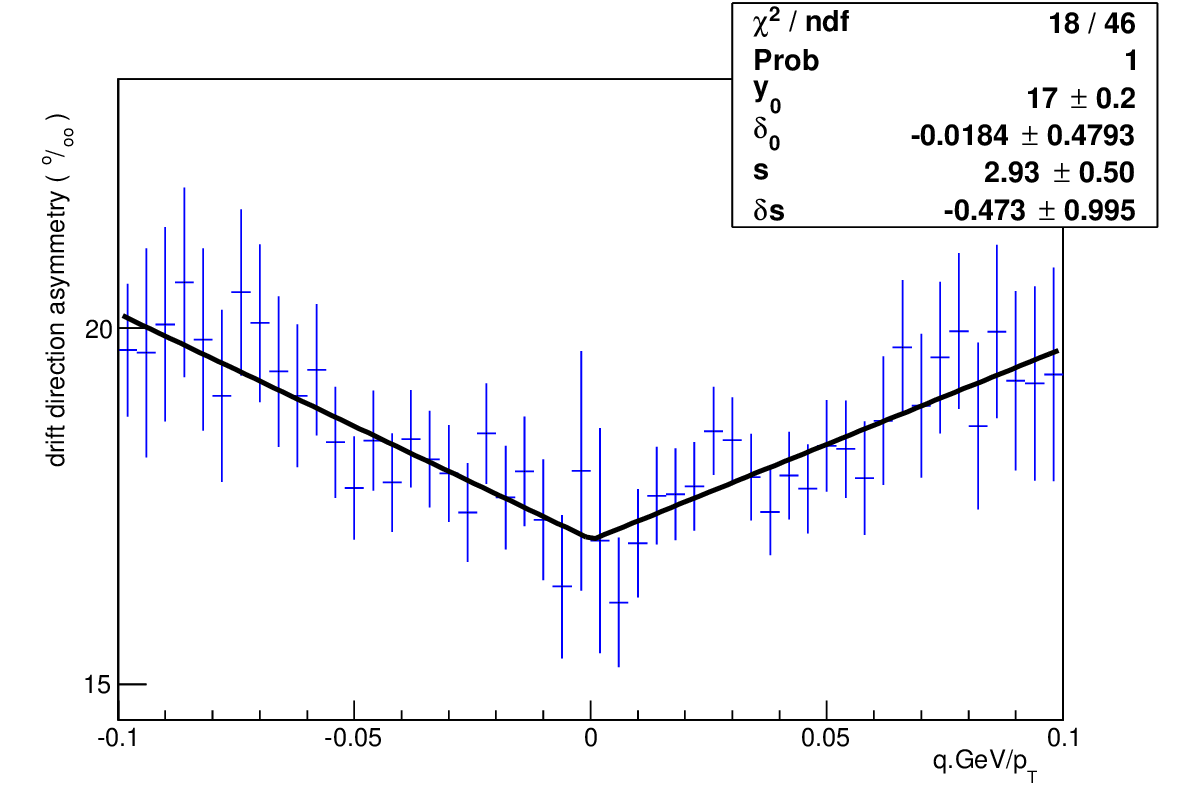}
\includegraphics[width=3.5in]{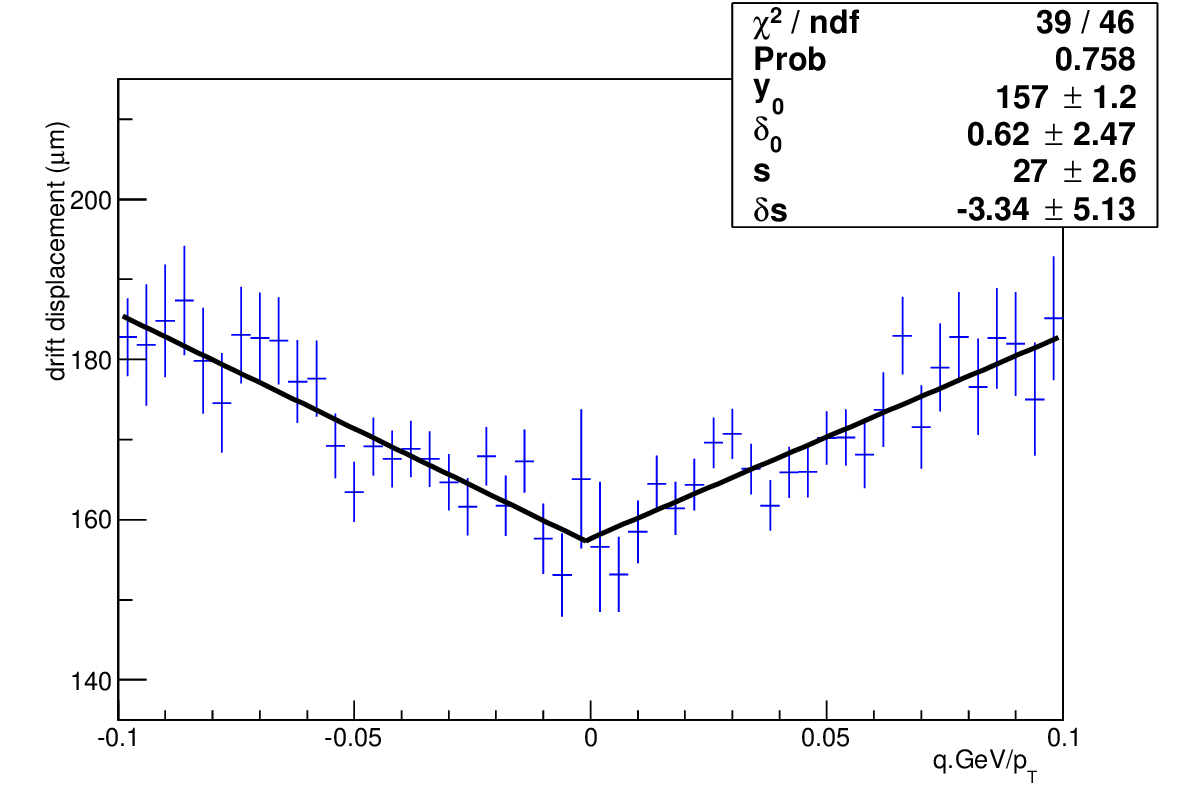}
\includegraphics[width=3.5in]{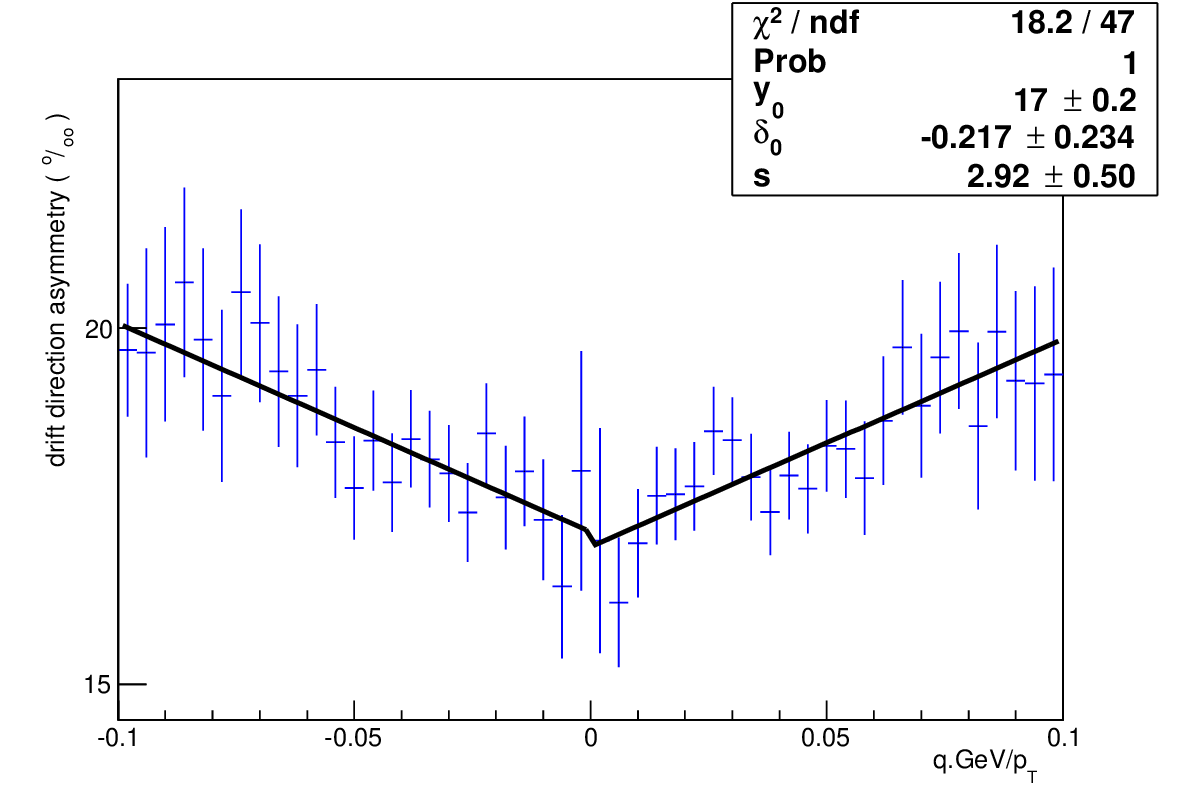}
\includegraphics[width=3.5in]{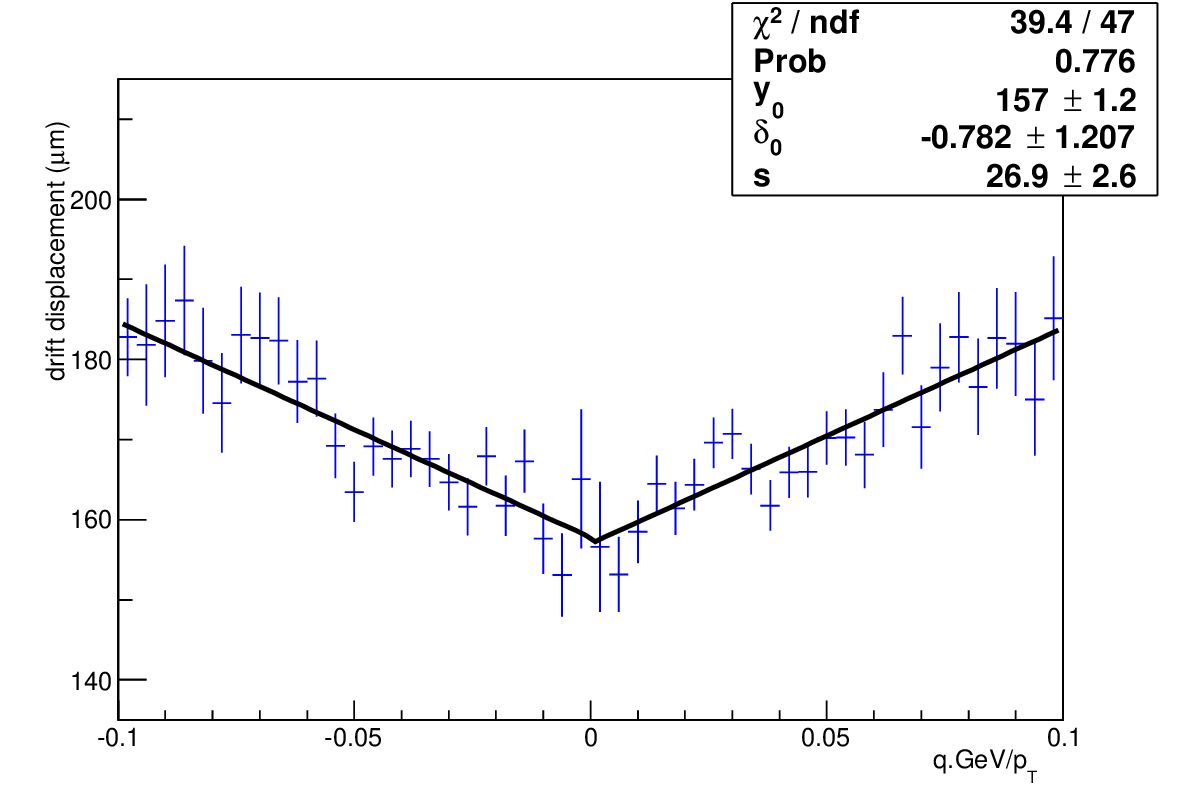}
\end{center}
\caption{Measurements of drift-cell properties using cosmic-ray data. (Left) The asymmetry between the fraction of hits drifting
 left versus right in each drift cell. (Right) The average drift displacement for all hits. (Top) The data are fit with the 
 functional form $y_0 + q\delta_0/2 + 10|c| (s + q \delta_s/2)$ where $c$ is in GeV$^{-1}$. (Bottom) The fit is repeated to the
 same data with the constraint $\delta_s = 0$.  }
\label{fig:lrAsym}
\end{figure}

 Of interest to this study is the dependence of these cell-level diagnostics on the track curvature. Again, we use the dicosmic helix
 to provide the most precise and accurate measurement of the curvature $c_{\rm d}$ of the cosmic-ray muons. The left-right asymmetry and 
 the average drift displacement for all cells is shown as a function of $c_{\rm d}$ in Fig.~\ref{fig:lrAsym}. In each case we fit for 
 four parameters using the following linear function of $|c_{\rm d}|$ (in GeV$^{-1}$), $y_0 + q\delta_0/2 + 10|c_{\rm d}| (s + q \delta_s/2)$, with
 an explicit dependence on charge included. Thus, $\delta_0$ captures any discontinuity in the $c \to 0^{\pm}$ limits and $\delta_s$
 captures any charge-dependence in the slope as one approaches these limits. 

 The charge-independent geometrical effects of the
 COT geometry are captured by the $y_0$ and $s$ parameters that describe these $p_T > 10$~GeV data. 
 The drift-direction asymmetry and the average drift displacement provide consistent descriptions of the cell drift
 where the maximum drift distance is 8.8~mm~\cite{cotNim}; multiplying this maximum distance by the directional asymmetry yields 
 predictions for the average displacement that are  consistent with the observed values of the latter. This consistency implies that the
 drift speeds are the same on both sides of the drift cell. 

The $\delta_0$ and $\delta_s$ values are consistent with zero and do not support the hypothesis of a $b_0q$ term. 
 Since these parameters are strongly anticorrelated, the fit is repeated with the constraint $\delta_s = 0$, 
 i.e. the slope is taken to be charge-independent. This improves the $\chi^2$/dof and the constraining power on $\delta_0$, 
 the parameter that tests for a discontinuity in the high-$p_T$ limit. We find no discontinuity within a precision of 1.2~$\mu$m. 

 In Appendix~\ref{appendixWireBias} a hypothetical charge-dependent alignment bias of 1~$\mu$m is used to estimate an $m_W$ bias of 4~ppb. Thus
 the above constraint on $\delta_0$ implies a bound of 5~ppb on any $m_W$ bias.
 
 A simulation of the COT geometry and drift cells can be used to explore potential sources of a $b_0q$ term; currently no physical 
 model for this term can be created. In Appendix~\ref{appendix:b0Model} we present an energy-loss model 
 as an illustration of an effect that generates this term, albeit with a vanishing value of the $b_0$ coefficient. 
 \subsection{Summary of singular response functions}
 \label{sec:singularSummary}
 \hspace*{0.06in}
A non-analytic response can be parameterized by terms of the form $(a_r + qb_r)|c|^r$ where $r$ is not a whole number. All such terms
 imply discontinuities (gaps in acceptance or anomalous charge drift) that are inconsistent
 with the construction of the COT. 
 We have demonstrated this fact by studying the hit efficiency and the drift displacement versus curvature, and
 showing that both observables vary extremely smoothly as $c \to 0^\pm$. The upper bound on any discontinuity in each observable corresponds to
 a systematic uncertainty on the momentum calibration of 5~ppb.

 We have shown that a non-analytic response is adequately captured by a term of the form $\delta c = b_0q$  to parameterize a discontinuity
 as $c \to 0^\pm$. We show in Appendix~\ref{appendix:b0Model} that synchrotron radiation induces this term with a momentum miscalibration of
 0.0001 ppb. 
 \section{Summary}
\hspace*{0.06in}
 We have discussed a curvature response model for the COT, the CDF experiment's drift chamber that operated during Run 2 of the Tevatron. In
 the preceding sections we have described how the model parameters $a_n$ and $ b_n$ ($n = 0,1,2,3$) capture the degrees of freedom relevant for
 calibrating the curvature response in the context of the $m_W$ measurement. The constraints on these parameters have been derived from cosmic-ray
 alignment data collected in situ with collider data, as well as the $E/p$ ratio of positrons and electrons from $W \to e \nu$ decays, and the
 $ J/\psi \to \mu \mu$ and $ \Upsilon \to \mu \mu$ data. These and other measurements over-constrain the model parameters and bound the
 systematic uncertainty on $m_W$, as summarized in Table~\ref{tab:parameterSummary}.
 \begin{table}[!ht]
\begin{center}
\begin{tabular}{|c|c|c|c|c|c|c|c|c|}
  \hline dataset  & $a_0$  & $b_0$ & $a_1$  & $b_1$  & $a_2$  & $b_2^\prime$ & $a_3$  & $b_3$  \\
  \hline cosmic & 0.2 & 0.005 & $\cal I$ & 0.3 & 0.008 & $\cal I$ & $\cal I$ & 0.003 \\
          rays & {\footnotesize (\ref{sec:mWZ2})} & {\footnotesize (\ref{sec:singularSummary})} & {\footnotesize (\ref{sec:deltaCm})} & {\footnotesize (\ref{sec:mWZ2})} & {\footnotesize (\ref{sec:mWZ2})} & {\footnotesize (\ref{sec:deltaCm})} & {\footnotesize (\ref{sec:deltaCm})} & {\footnotesize (\ref{sec:mWZ2})} \\
  \hline $J/\psi, \Upsilon$ & $\cal N$ & $\cal N$ & 25 & $\cal N$ & $\cal N$ & 1 & 12 & 4  \\
  $\to \mu \mu$ & {\footnotesize (\ref{sec:mass1})} & {\footnotesize (\ref{sec:mass1})} & {\footnotesize (\ref{sec:intro})} &  {\footnotesize (\ref{sec:mass1})}  &  {\footnotesize (\ref{sec:mass1})}  & {\footnotesize (\ref{sec:mass1})} & {\footnotesize (\ref{sec:a3})} & {\footnotesize (\ref{sec:b3})} \\
  \hline $\Delta_W^\mp$  & 0.04  & $\cal I$ & $\cal I$ & 0.04 & $\cal N$ & $\cal I$ & $\cal I$ & $\cal N$ \\
   {\footnotesize $(W \to \ell \nu)$}  & {\footnotesize (\ref{sec:wmass2})} & {\footnotesize (\ref{sec:wmass1})} & {\footnotesize (\ref{sec:wmass1})} & {\footnotesize (\ref{sec:wmass2})} & {\footnotesize (\ref{sec:wmass2})} & {\footnotesize (\ref{sec:wmass1})} & {\footnotesize (\ref{sec:wmass1})} & {\footnotesize (\ref{sec:wmass2})} \\
  \hline $\Delta_{pe}$  & 0.002 & $\cal I$ & $\cal I$ & 0.002 & $\cal N$ & $\cal I$ & $\cal I$ & $\cal N$ \\
   {\footnotesize $(W \to e \nu)$}  & {\footnotesize (\ref{sec:wmass2})} & {\footnotesize (\ref{sec:eop})} & {\footnotesize (\ref{sec:eop})} & {\footnotesize (\ref{sec:wmass2})} & {\footnotesize (\ref{sec:wmass2})} & {\footnotesize (\ref{sec:eop})} & {\footnotesize (\ref{sec:eop})} & {\footnotesize (\ref{sec:wmass2})} \\
\hline
\end{tabular}
\end{center}
\vskip -0.1in
\caption{The constraints obtained from independent datasets  on  the $a_n$ and $b_n$ parameters 
   of the curvature response model (Eq.~\ref{eq:analytic} and Sec.~\ref{sec:singularImpact}). The constraints   are presented in terms of the 
  corresponding uncertainty (in ppm) on the $W$ boson mass.  
  Note that $b_2^\prime \equiv b_2 + \epsilon$ where $\epsilon$  is the energy loss incurred by a muon as it traverses the tracker. The 
  half-difference of $\langle E/p \rangle$ between positrons and electrons in $W \to e \nu$ decays
   is denoted by $\Delta_{pe}$ as shown in Eq.~\ref{eq:deltaPE}.
  The fractional half-difference in $m_W$ between positive and negative leptons is denoted by $\Delta_W^\mp$ as shown in Eq.~\ref{eq:deltaW}.
  The numbers in parentheses refer to the section of this document where the corresponding details are provided. The $\cal I$ entries
 indicate parameters that are invisible to the corresponding observables. The $\cal N$ entries
 indicate parameters that are not constrained by the corresponding observables. Note that the constraint on $a_1$ from the $J/\psi \to \mu \mu$
 and $\Upsilon \to \mu \mu$ data already takes into account the effects of the $b_2^\prime$, $a_3$ and $b_3$ parameters in the quoted 25~ppm
  uncertainty~\cite{CDF2022}. } 
\label{tab:parameterSummary}
\end{table}

 Table~\ref{tab:parameterSummary} reiterates the conclusion from Sec.~\ref{sec:wmassTot} that the analytic model of Eq.~\ref{eq:analytic} is highly
 constrained by the combination of the datasets. In particular, the analytic parameters that contribute the largest momentum calibration uncertainty
 on $m_W$ are already accounted for in the analysis of the $J/\psi$ and $\Upsilon$ data. All other parameters contribute negligibly to the uncertainty
 on $m_W$.

 We also considered the impact of all  parameters on the $m_{J/\psi}$ and $m_\Upsilon$ analyses.
  Table~\ref{tab:mesonSummary} summarizes the constraints
 from cosmic-ray data and their impact on the momentum scale parameter $a_1$ extracted
 from the $J/\psi$ and $\Upsilon$ data. Table~\ref{tab:mesonSummary} reiterates the conclusion from Sec.~\ref{sec:wmassTot} that the
  momentum calibration is robust with respect to analytic deformations of the COT. 
 \begin{table}[!ht]
\begin{center}
\begin{tabular}{|c|c|c|c|c|c|c|c|c|}
  \hline analysis  & $a_0$  & $b_1$  & $a_2$  & $b_3$ & Sec. & $b_0$ & Sec.\\
  \hline $m_{J/\psi}$ & 0.001 & 0.3 & 0.8  & 5 & {\footnotesize \ref{sec:mJ2}} & $<0.001$ & {\footnotesize \ref{sec:singularSummary}} \\
  \hline $m_\Upsilon$ & 0.003 & 0.3 & 0.4 & 1 & {\footnotesize \ref{sec:mUpsilon2}} & $<0.001$ & {\footnotesize \ref{sec:singularSummary}} \\
\hline
\end{tabular}
\end{center}
\vskip -0.1in
\caption{The constraints obtained from cosmic-ray data  on a subset of the $a_n$ and $b_n$ parameters
   of the curvature response model (Eq.~\ref{eq:analytic} and Sec.~\ref{sec:singularImpact}). The constraints   are presented in terms of the
   corresponding uncertainty (in ppm) on the $m_{J/\psi}$ and $m_\Upsilon$ analyses; these constraints can be interpreted as the corresponding
    uncertainties on the $a_1$ extracted from the $J/\psi$ and $\Upsilon$ data. 
  The remaining parameters, $b_2^\prime$ and $a_3$ have already been accounted for in the quoted 25~ppm uncertainty on $a_1$~\cite{CDF2022}. 
  The columns labelled ``Sec.'' refer to the section of this document where the corresponding details on the preceding columns are provided. }
\label{tab:mesonSummary}
\end{table}

 Finally, a detailed analysis of non-analytic response functions has been performed. It is shown that such functions imply a discontinuity
 as $c \to 0^\pm$ which is adequately captured by a term of the form $\delta c = b_0 q$. Cosmic-ray data are used to study the smoothness of
 the COT efficiency and drift displacement in the $c \to 0^\pm$ limit and the systematic uncertainty on $m_W$ is constrained to 5~ppb due
 to any discontinuity. This is noted in Tables~\ref{tab:parameterSummary} and~\ref{tab:mesonSummary}. Cosmic-ray data also show that the COT
 efficiency is extremely stable in time; the superlayer inefficiency is consistent with a constant value of $(520 \pm 30)$~ppm over its 
  entire 10-year operation. 

  Having considered a general curvature response model, we find that all parameters are tightly constrained by control samples of in-situ cosmic-ray
  data, $J/\psi$ and $\Upsilon$ data in the dimuon channel, and the charge-asymmetry of lepton $p_T$ measurements. Uncertainties due to
   all other parameters are found to be negligible in comparison with the quoted uncertainty on the momentum calibration~\cite{CDF2022}. 
\section{Conclusions}
 \hspace*{0.06in}
We have studied a parametric model for the curvature response of the CDF experiment's drift chamber, which provides the momentum 
measurement of charged particles. This investigation includes terms that are analytic in curvature as well as terms that 
 capture non-analytic behavior in the limit of vanishing curvature. The study includes in situ cosmic-ray data recorded by the CDF
 detector during collider operation, as well as information from the publications describing the 
 measurement of the $W$ boson mass. 

We find that the analytic terms in the most general ansatz have well-defined physical interpretations.
 They are either constrained within the published uncertainties or contribute negligibly to the uncertainty on the momentum calibration. 
Furthermore, the analysis presented here on the basis of first principles shows how parameter 
 uncertainties can be controlled and understood  
 without recourse to black-box methodology such as machine learning or high-dimensional fitting. 
 
 We find no realistic model that would generate significant non-analytic terms without violating the principles of operation of the 
 drift chamber or the physics of particle interactions. Non-analytic terms are further constrained by analysis of the cosmic-ray data. 

 Without loss of generality, uncertainties due to
 all other parameters are found to be vanishing in comparison with the quoted uncertainty on the momentum calibration in the CDF $m_W$
  measurement~\cite{CDF2022}, demonstrating the explanability and robustness of these procedures.
 
 This study provides a framework for further investigation of this drift chamber and the analyticity of the curvature response 
 of tracking detectors in general. The topic is of relevance to high-precision measurements of observables such as the $W$ boson mass
 and the weak mixing angle where accuracy of particle tracking is important, at future fixed-target and collider experiments.
 A gaseous detector with a single unfragmented active volume can be calibrated at the level of accuracy required to meet the goals of those
  experiments. \\

\noindent  {\bf Acknowledgements} \\
 \hspace*{0.06in}
We thank our colleagues on the CDF experiment at the Fermi National Accelerator Laboratory. We thank Josh Bendavid, Maarten Boonekamp, Daniel Froidevaux, Paul Grannis, Hugh Montgomery, Jan Stark
and Charles Young for helpful discussions. We are grateful to Fermilab for providing computing resources and we acknowledge the support of the U.S. Department of Energy.

\appendix
\section{Correspondence of curvature response models}
\label{appendixModelCorrespondence}
\hspace*{0.06in}
The curvature response model presented here generalizes Eq.~23 of~\cite{CDF2firstPRD},
$$ \delta c = \epsilon_1 + \epsilon_2 c + \epsilon_3 c^2 + \epsilon_4 c^3$$ with the correspondence $\epsilon_{n+1} \leftrightarrow a_n$ as used in
 Eq.~\ref{eq:analytic} of this
 document. In this work we have extended the model to include the $b_n q$ terms.

 As noted in Sec.~\ref{sec:mass1}, the parameter $b_2$ and
the ionization energy loss appear in combination as a sum when these models are applied to outgoing particles. The ionization energy loss
is explicitly considered as a tunable parameter in all CDF publications, i.e., Eq.~27 of ~\cite{CDF2firstPRD}, Sec.~VII.B.3 of~\cite{CDF2014}  and
Sec.~VI.A of~\cite{CDF2022}. Therefore the inclusion of $b_2$ in this work does not materially augment the model; the new parameters
we have included are $b_0$, $b_1$ and $b_3$. More importantly, we have shown that, with the inclusion of these parameters,
the model is fully generalized for $p_T > 2$~GeV.

In previous CDF II publications, models were introduced to capture the spatial variation of $a_0$ after the cosmic-ray alignment; in particular, 
Eq.~24 of~\cite{CDF2firstPRD} and Eq.~14 of~\cite{CDF2014} parameterize the dependence of our $a_0$ parameter on the azimuthal and polar angles. 
After those earlier CDF publications, the cosmic-ray alignment of the COT was considerably improved~\cite{cosmicAlignment}, as was the determination 
of the beam coordinates, and no further azimuthal variation was observed (as demonstrated in~\cite{cosmicAlignment}).
 As a result, most of the parameters of Eq.~24 of~\cite{CDF2firstPRD} and 
Eq.~14 of~\cite{CDF2014} were no longer needed for the latest CDF publication of $m_W$ (Eq.~S4 of~\cite{CDF2022}).

As explained in Appendices~\ref{appendixCRalign} and~\ref{appendixEoPalign}, a small dependence on polar angle remains after the cosmic-ray alignment
due to the residual inaccuracy of the wire-shape model. The same quadratic parameterization is used in all three CDF II publications to correct for
this residual misalignment. The nomenclature $A, B, C$ is used in Eq.~15 of~\cite{CDF2014} and in Eq.~S4 of~\cite{CDF2022} (where the values of
$A$ and $C$ are displayed), while the correspondence $A \to a_0$, $B \to a_1$, $C \to a_2$ matches Eq.~24 of~\cite{CDF2firstPRD}, for the
parameters of the quadratic polar-angle dependence.

Thus, the symbols used in the earlier parameterizations to describe spatial variation of the alignment 
  have no correspondence with the $a_n$, $b_n$ notation used in this work to  describe curvature-dependent effects. 
\section{Global COT alignment}
\hspace*{0.06in}
The COT is well-aligned with the solenoid axis and accurately aligned with the beam axis. 
\subsection{Mutual alignment of the COT and the beam axis}
\label{appendixBeamCentering}
\hspace*{0.06in}
The consistency of the COT coordinate system and the location of the beam axis is important in order to eliminate bias when the 
 COT tracks are constrained to the beam coordinates. This consistency is ensured using a large collection of COT tracks of high quality.
 Their two-dimensional impact parameter vectors with respect to the beam axis (as recorded in the database) is used to compute any shift 
 between the COT coordinate system and the beam axis. The COT is then shifted in the track reconstruction software and the data are 
 re-reconstructed to confirm that the mutual alignment is now perfect (at the submicron level) since there is no shortage of data for 
 this procedure. Convergence is achieved typically in one iteration and a maximum of two iterations.  

This alignment is performed in separate run blocks by first identifying relative shifts between the beam axis and the COT coordinate
  system  over the entire 10 years of operation. The run blocks integrate data between occurrences of significant shifts. The procedure
 ensures submicron-level mutual alignment over the full collider dataset.  

One of the benefits of this accurate mutual alignment is that prompt COT tracks can be constrained to include the beam spot of 
 transverse size $\sim 30~\mu$m in the track fit, which improves the curvature (and thereby momentum) resolution considerably. There
 is no benefit from including silicon vertex detector information in the track fit, because the incremental improvement in momentum
 resolution is marginal and the COT track resolution is already good enough to not limit any aspect of the $m_W$ 
 measurement~\cite{CDF2022}.
\subsection{Mutual alignment of the COT and the solenoid}
\label{appendixSolenoid}
\hspace*{0.06in}
The COT is mounted on the inside of the solenoid support structure. The time-of-flight detector is installed between the COT and the solenoid, leaving free space of $\cal O$(1~mm) between
the detectors. This tolerance limits the relative displacement or tilt of the COT with respect to the solenoid axis. 

The static magnetic field inside the solenoid varies smoothly according to Laplace’s equation. 
The magnetic field has circular symmetry and cannot have a first-order transverse gradient near its axis. A small transverse displacement between the COT and solenoid axes cannot
 change the field near the axis at first order. Away from the axis any change in the field averages over azimuth to zero at first order. 
 The longitudinal component of the magnetic field changes by the cosine of a tilt angle; as the latter is limited to $\cal O$(1~mm/1~m~$\approx 1$\permil), $B_z$ is affected at $\cal O$(1~ppm) 
 which  is negligible. 

 Miscalibration of the magnetic field can only scale the track $p_T$ by a smooth spatial function which precludes singular behavior in the $c \to 0$ limit, i.e., it cannot introduce singular
 terms in Eq.~\ref{eq:analytic}. Furthermore, $a_0$, $b_1$, $a_2$ and $b_3$ ($B$ terms in Eq.~\ref{eq:B}) cannot be induced because the magnetic field affects both charges symmetrically.
  Any induced $A$ terms ($a_1$, $b_2$ and $a_3$ in Eq.~\ref{eq:A}) are calibrated using the $J/\psi$ and $\Upsilon$ data
  (Sec.~\ref{sec:wmassTot}) modulo the small non-uniformity of the magnetic field which is not averaged identically by the $J/\psi$, $\Upsilon$, $W$ and $Z$ boson decays. The field non-uniformity versus polar angle 
   (a small fringe-field miscalculation) is calibrated using the $J/\psi$ data in the CDF $m_W$ analyses~\cite{CDF2022,CDF2014,CDF2firstPRD} with an
  uncertainty due to differences between the spatial distributions of the decay muons in the various signal  samples. 
\section{Ionization energy loss for cosmic rays}
\label{appendixEloss}
\hspace*{0.06in}
For a transverse energy loss $\epsilon$,  the measured curvature is $c^{\rm measured} = q / (p_T - t \epsilon)$ where $t = \texttt{+}1(\texttt{-}1)$ for outgoing (incoming) particles. Defining $\zeta \equiv \epsilon / p_T = \epsilon q c$, 
\begin{eqnarray}
\delta c \equiv  c^{\rm measured} -c & = & \frac{q}{p_T - t \epsilon} - \frac{q}{p_T} = c (\frac{1}{1 - t \zeta} - 1) = c \frac{t \zeta}{1 - t \zeta} \nonumber \\
& \simeq & c t \zeta (1 + t \zeta) = c(t \zeta + \zeta^2) = c(t q \epsilon c + \epsilon^2 c^2) \nonumber
\end{eqnarray}
Therefore the energy loss induces the correction terms $\epsilon c^2(t q + \epsilon c)$ in the measured curvature of the incoming and 
outgoing muons respectively.

Hard scattering from the drift chamber wires introduce small discontinuities in the curvature of the tracks. The impact of hard scattering may be estimated using the Rutherford
  formula for the differential scattering cross section, 
\begin{eqnarray}
  \frac{d\sigma}{d\Omega} = \left(\frac{Z\alpha}{4E_\mu \sin^2(\theta/2)}\right)^2 \approx \left(\frac{Z\alpha}{E_\mu \theta^2}\right)^2
\end{eqnarray}
for a nucleus of atomic number $Z$, lepton energy $E_\mu$ and small scattering angle $\theta$, where $\alpha = \frac{e^2}{4\pi} = 137^{-1}$ is the fine-structure
 constant for lepton charge $e$.

A 40~GeV lepton traversing a 1.4~T magnetic field produces a curvature of $53 \times 10^{-6}$/cm and a maximum sagitta of $\approx 9$~mm in the COT. This curvature is mimicked  
  by a hard-scattering angle of 13~mrad at the middle radius of the COT. Since any hard scattering will create a deflection that is uncorrelated with the sign of the
   lepton charge and its
   track curvature, the deflection angle will not bias the true curvature but will cause fluctuations around the true curvature.
   A deflection of 0.4~mrad, when projected on the transverse plane by a factor of $1/\sqrt{2}$,  will contribute a resolution of $\approx$2\% on the curvature. The
   intrinsic curvature resolution of the COT~\footnote{The resolution of beam-constrained COT tracks is $\delta({\rm GeV}/p_T) = 0.5$\permil ~\cite{CDF2firstPRD}.}
   at 40~GeV is 2\%, therefore the impact of a 0.4~mrad deflection is similar to the intrinsic resolution.

The probability of a hard-scatter deflection larger than 0.4~mrad can be conservatively estimated by integrating the Rutherford scattering differential
   cross section, yielding the cross section 
\begin{eqnarray}
  \sigma  \approx \pi\left(\frac{Z\alpha}{E_\mu \theta}\right)^2 = \pi\left(\frac{74/137}{40,000 ~ {\rm MeV} \times 0.0004}\right)^2 =
   \pi\left(\frac{74/(137 \times 16)}{ {\rm MeV} }\right)^2 \approx \pi(6.7~{\rm fm})^2
\end{eqnarray}
for a 40~GeV lepton scattering off a tungsten nucleus. The probability $p$ of a hard scatter 
 is given by $p = \sigma n d$ where $n$ is the number density of scattering centers (atoms) and $d$ is the thickness of the scattering layer. For a tungsten wire of
 40~$\mu$m diameter~\cite{cotNim} in the COT, $n = 6.4 \times 10^{22}/$cm$^3$ yields 
 $p = \pi(6.7 \cdot 10^{-13} ~ {\rm cm})^2 \times (6.4 \cdot 10^{22} ~ {\rm cm}^{-3}) \times (40 \cdot 10^{-4} ~ {\rm cm}) = 0.36$\permil. 

Each sense wire sits in the middle of a drift region of width $2 \times 8.8$~mm. The probability of a particle hitting a specific wire  is
  (40~$\mu$m)/(17.6~mm) = 2.3\permil. The corresponding poisson probability of the hard scatter is 0.36\permil\ $\times$ 2.3\permil\ $\approx 1$~ppm. Summing over the wires in the radial direction, the total
   probability of a hard scatter with $\theta > 0.4$~mrad in \textit{any} wire is 0.2\permil. 

Repeating this analysis for different $\theta$-thresholds leads to the same conclusion. For example, lowering the threshold by $10\times$ will increase
  the rate $100\times$ to 2\%, but these smaller-angle scatters will contribute to the intrinsic resolution in quadrature and increase the latter by only 1\% of itself;
  the net effect will again be 0.2\permil\ on the resolution. If we raise the threshold by $10\times$, these large-angle scatters can affect the momentum measurement by
  20\%; however their rate of 2~ppm is lower than misidentification backgrounds from other sources by three orders of magnitude~\cite{CDF2022,CDF2firstPRD,CDF2014}. 
  Finally, we note that coherent elastic scattering off the nucleus, per the Rutherford formula, dominates over incoherent, inelastic scattering; thus the latter can also be
 ignored. 

These estimates show that the probability of a hard scatter is so small that it modifies the intrinsic
   resolution at less than the permille level. Furthermore, the resolution is modeled using
   hit residuals measured for the muon tracks in $Z \to \mu \mu$ data~\cite{CDF2022}; thus the model already subsumes the tiny contribution from hard scattering. The 
   observed non-Gaussian tails in the hit residuals, which are caused by complexity of pattern recognition,
      are incorporated in the model~\cite{CDF2022}. The small non-Gaussian (power-law) tail due to hard scattering is also subsumed. 

  This is the reason that the COT track fit
  is a simple $\chi^2$-minimizing fit. More sophisticated methods such as the Kalman Filter are unnecessary for a tracking detector that is as transparent as the COT. Furthermore, the simplicity
  of the simple $\chi^2$-minimizing fit provides the advantage that COT tracking is easy to simulate from first principles, which aids in its accurate calibration. This is one of the motivations
  for building a transparent gaseous tracker such as a drift chamber or a time-projection chamber at a future electron-positron collider, where the tracker must be
 calibrated with extreme accuracy for measuring precision electroweak and Higgs observables. 

 Pions, kaons and protons undergo additional hard scatters in the COT due to hadronic interactions. The thickness of tungsten traversed on average
   through the entire COT is $18~\mu$m or 0.3\permil\ of a nuclear collision length. This rather small probability of a hadronic
  interaction explains why the same simple $\chi^2$-minimizing fit in COT tracking is used for all particles, leptons and hadrons. 
\section{Spatial uniformity of the COT}
\label{appendixUniformity}
\hspace*{0.06in}
Figures~\ref{deltaPlus} and~\ref{deltaPlusReduced} show the post-alignment constraints from cosmic-ray data
 on the $a_0$, $b_1$, $a_2$ and $b_3$ parameters; only $b_3$ is statistically significant. 
Tables~\ref{tab:parameterSummary} and~\ref{tab:mesonSummary} show that $b_3$ has no impact on the $m_W$, $m_{J/\psi}$ and $m_\Upsilon$ analyses. Hence
the spatial variation of $b_1$, $a_2$ and $b_3$ is moot.

The  $a_0$ term is directly induced by misalignments. This is clear from Eq.~\ref{eq:analytic}; if the cosmic-ray alignment were performed with
zero magnetic field\footnote{In order for the cosmic-ray alignment to be applicable to physics measurements, the cosmic-ray sample must be collected
in situ with collider data at full solenoidal magnetic field. The COT drift cells are designed to operate at the corresponding Lorentz angle. The
drift model would not be appropriate for zero magnetic field and the alignment derived with a different drift model may not be applicable for full-field
 operation.},
 the cosmic-ray trajectories would be straight lines with zero curvature and only $a_0$ would be sensitive to misalignments.
 It is therefore natural to associate with the $a_0$ parameter
 the  \textit{a priori} dependence of the alignment on the azimuthal $(\phi)$ and polar $(\theta)$ angles. 
 The zero-field situation is effectively mimicked by reweighting positive cosmic rays to obtain the sample average $\langle c  \rangle = 0$. 
 
 Formally, the interplay between the $a_n$ and $b_n$ coefficients in the alignment procedure can be understood as follows. The positions of the
 supercells are adjusted to minimize the residuals with respect to the dicosmic fit~\cite{cosmicAlignment}; as a consequence, the sample
 average $\langle \Delta_c^+ \rangle$ is minimized. Eq.~\ref{deltaPlus} implies that the quantity 
 $[ a_0 + b_1 \langle |c| \rangle + a_2 \langle |c|^2 \rangle +  b_3 \langle |c|^3 \rangle ]$ is minimized. We ignore the statistically-insignificant
 coefficients $b_1$ and $a_2$. The $|c|$ distribution is approximately a Gaussian with mean 35~TeV$^{-1}$ and RMS 25~TeV$^{-1}$, thus 
 $\langle |c|^3 \rangle \sim 110,000$~TeV$^{-3}$. Using $b_3 \sim 0.022$~GeV$^{2}$ from Fig.~\ref{deltaPlusReduced} yields
 $b_3 \langle |c|^3 \rangle \sim 2.5$~PeV$^{-1}$. Thus the effect of $b_3$ is much smaller than the statistical uncertainty of 10~PeV$^{-1}$
 on $a_0$ and can be ignored; spatial uniformity needs to be  addressed in the context of $a_0$ only.  
\subsection{Benefits of symmetry and continuity}
\hspace*{0.06in}
The simplicity of the COT geometry and its azimuthal (Fig.~\ref{fig:COTquad}) 
and longitudinal symmetries limit the \textit{a priori} spatial variation of the $a_0$ parameter. Table~\ref{tab:alignmentUniformity}
shows that this variability is similar to the intrinsic resolution.

The longitudinal symmetry is broken by the deviation of the wires' shape
from a straight line due to electrostatic deflection and gravitational sag. Nevertheless, only a handful of parameters are required to describe
the shape analytically; it must be a smooth, quadratic function of $|z|$ at static equilibrium and it must be 
a smooth, sinusoidal function of azimuth due to the interplay between the azimuthal symmetry of the COT end plates and the direction of
 gravity~\cite{cotNim}.

These simplifying features distinguish a
drift chamber with a single active volume from the complicated geometry of a silicon tracker which is a patchwork of planar tiles. 

\subsection{Benefits of cosmic-ray alignment} 
\label{appendixCRalign}
\hspace*{0.06in}
The \textit{a priori} azimuthal 
variation  is suppressed by two orders of magnitude by the cosmic-ray alignment (see Table~\ref{tab:alignmentUniformity}). Importantly, the remaining
 variation is consistent with statistical scatter~\cite{cotNim}.  Further studies can therefore be performed inclusively over azimuth.
 
  \begin{table}[!ht]
  \small 
\begin{center}
  \begin{tabular}{|l|c|c|c|c|c|c|c|}
  \hline \multicolumn{1}{|c}{pull} & \multicolumn{2}{|c|}{azimuth} & \multicolumn{2}{|c|}{$z_0$} & \multicolumn{2}{|c|}{$\cot \theta$} & RMS \\
\multicolumn{1}{|c|}{observable}     & pre-  & post-  & pre-  & post-  & pre-  & post-  &  \\
  \hline $\Delta^+_c$ (PeV$^{-1}$) & $[-1200,0]$  & $\pm 20$  & $[-620,-580]$  & $\pm 40$ & $[-800,-500]$  & $\pm 30$ & 930 \\
  \hline $\Delta \phi_0$ ($\mu$rad) & $[-1400,0]$  & $\pm 5$  & $[-650,-500]$  & $\pm 15$ & $[-600,-500]$  & $\pm 10$ & 840 \\
  \hline $\Delta d_0$ ($\mu$m) & $[-400,100]$  & $\pm 2$  & $[-180,-160]$  & $\pm 8$ & $[-230,-150]$  & $\pm 4$ & 310 \\
\hline
  \hline $\Delta \cot \theta$ ($10^{-6}$) & $[-4000,3000]$  & $\pm 40$ & $[-2500,-500]$  & $\pm 150$ & $[-1600,-1200]$  & $\pm 50$ & 4300 \\
  \hline $\Delta z_0$ ($\mu$m) & $[-3000,3000]$  & $\pm 40$ & $[-350,-250]$  & $\pm 30$ & $[-400,-300]$  & $\pm 40$ & 3800 \\
\hline
  \hline $\Delta t_0$ (ps) & $[-150,300]$  & $8 \pm 8$ & $[0,20]$  & $[0,20]$ & $[0,30]$  & $[0,30]$ & 430 \\
\hline
  \end{tabular}
\end{center}
\vskip -0.1in
\caption{A summary of the variability displayed by the pulls as a function of azimuth, $z_0$ and polar angle respectively (reproduced from 
Figs.~20-25 of~\cite{cosmicAlignment}). Each pull is defined by the comparison of the observable between the incoming and outgoing legs of the cosmic ray.
    Note that $\Delta^+_c$ (Eq.~\ref{deltaPlus}) is defined as half of the pull defined in~\cite{cosmicAlignment}.
 The $\phi_0$, $d_0$ and $z_0$ ($t_0$) observables denote the azimuthal angle, the impact parameter and the $z$-position (time of occurrence) 
  respectively of the tracks at the point of closest approach to the beam axis.
 ``pre-'' and ``post-'' refer to pre-alignment and post-alignment with the cosmic-ray procedure, and ``RMS'' refers to the RMS of inclusive
  distribution of the pull after alignment, i.e., the intrinsic resolution. The first three (next two) rows of observables describe the track's
  trajectory in the transverse plane (longitudinal view). The 8~ps bias in $t_0$ translates to an immaterial 0.4~$\mu$m effect using the drift speed
   of 50~$\mu$m/ns in the COT. 
}
\label{tab:alignmentUniformity}
\end{table}

  The \textit{a priori} variability with respect to
   the longitudinal variables $z_0$ (position of the track along the $z$-axis) and polar angle is even smaller
  than the azimuthal variability, reflecting a good understanding of the wire deflections due to electrostatic and gravitational forces. To achieve
   the ultimate goal of the accuracy of the alignment, the radial dependence of the wire shape is investigated in~\cite{cosmicAlignment}, because 
   wire tensions likely cause the end plates to bend inward with a radius-dependent displacement. With additional tuning of the
   wires' shape to accommodate this hypothesis, the variability of the pulls with respect to $z_0$ and $\cot \theta$ is suppressed by 1-2 orders of
    magnitude, as   shown in~\cite{cosmicAlignment} and summarized in Table~\ref{tab:alignmentUniformity}. The pulls are largest 
    near the edges of the phase space in $z_0$ and $\cot \theta$, both before and after this tuning. This is
     consistent with the hypothesis that inaccuracies
    in the wire shape are the root cause since the shape changes most rapidly at large $|z|$~\cite{cotNim}. 

    The post-alignment variability with respect to $z_0$ and $\cot \theta$ is only a factor of 1.5-2 larger than statistical scatter. Any systematic
    impact of averaging over this variation is further mitigated because studies with cosmic rays are performed in the same phase space
    as collider data: $|z_0| < 60$~cm and the fiducial acceptance of all COT superlayers. Thus the average $a_0$ for cosmic-ray data matches the
    collider data much more closely than this residual variability.

    \subsection {Benefit of COT-beam alignment}
    \hspace*{0.06in}
    Once the COT is internally well-aligned, accuracy of the beam coordinates is important. 
    Upon constraining the COT track to the beam axis, a $\sin \phi$ modulation will be induced if the beam-axis coordinates are incorrect in
the COT coordinate system. Eq.~24 of~\cite{CDF2firstPRD} and Eq.~14 of~\cite{CDF2014} include this term in the model.

In Appendix~\ref{appendixWireBias} we show that a 1~$\mu$m misalignment of the beam axis
induces a 5~PeV$^{-1}$ bias in the beam-constrained curvature; as mentioned in Appendix~\ref{appendixBeamCentering}, the beam axis has been pinned
down over the entire running period to better than this. Averaging a sinusoidal modulation over azimuth suppresses any residual bias by at least
another order of magnitude. Thus, any $\phi$-asymmetry due to beam-constraining will be far smaller than the 10~PeV$^{-1}$ estimate used
 for Tables~\ref{tab:parameterSummary} and~\ref{tab:mesonSummary}, and subsequent studies can be performed inclusively over azimuth. 

    \subsection {Ultimate $\Delta_{pe}$ tune}
\label{appendixEoPalign}
    \hspace*{0.06in}
    The $a_0$ parameter features prominently in the $\Delta_{pe}$ observable (Eq.~\ref{eq:deltaPE} and Sec.~\ref{sec:mWZ2}) and has been used
    to derive a final alignment correction as a quadratic function of $\cot \theta$, the relevant physical variable. The average over the
    $z_0$-profile of the beam luminous region applies consistently for all collider data, and azimuthal variation has already been eliminated.
    The azimuthal dependence modeled in Eq.~24 of~\cite{CDF2firstPRD} and Eq.~14 of~\cite{CDF2014} was not required in Eq.~S4 of~\cite{CDF2022},
    where no systematic azimuthal variation of $\Delta_{pe}$ was found. The parameters capturing the dependence on $\cot \theta$ have the same
    meaning in all three models, with a change of notation from~\cite{CDF2firstPRD} to~\cite{CDF2014,CDF2022} as mentioned 
     in Appendix~\ref{appendixModelCorrespondence}. 
    
    The coefficient linear in $\cot \theta$ is the only one that induces a non-negligible systematic uncertainty on $m_W$ because it couples to the charge
    asymmetry in $W$-boson production at the Tevatron. It is strongly constrained by the cosmic-ray data which span the COT in the $z$-direction
    and pin down the relative rotation between the end plates, i.e., the end-to-end twist of the COT. The twist degree of freedom for each COT
    supercell is explicitly measured in the cosmic-ray alignment procedure~\cite{cosmicAlignment}, which explains why the $\Delta_{pe}$ tune
    finds the twist coefficient to be $(0 \pm 4_{\rm stat})$~PeV$^{-1}$ (Eq.~S4 of~\cite{CDF2022}). The remaining two parabolic coefficients of the
     $\Delta_{pe}$ tune have non-zero values that are consistent with the imperfections
    in the wire-shape corrections derived from cosmic-ray data (Fig.~25 of~\cite{cosmicAlignment}) as noted in
    Table~\ref{tab:alignmentUniformity}.  The    internal consistency of the alignment over independent control samples of data gives confidence
    in robustness of the procedure.
    
    The twist coefficient induces a small     systematic uncertainty of 0.8~MeV on $m_W$, while the remaining two parabolic coefficients have
     negligible impact~\cite{CDF2022}. 
\section{Analysis of $\Delta^+_c$ and $\Delta^-_c$}
\label{appendixB}
\hspace*{0.06in}
In Sec.~\ref{sec:cosmics} we used the fitted dicosmic helix' curvature $c_{\rm d}$ as a proxy for the true curvature $c$ because the resolution 
 of $c_{\rm d}$ is a factor of 12 better than the curvature resolution of the separate measurements of the incoming and outgoing
 muons. We check whether this substitution
 allows us to be sensitive to the systematic bias $\delta c$ of Eq.~\ref{eq:analytic}, since the same bias may be present in $c_{\rm d}$, 
 i.e. $c_{\rm d} = c + \delta c_{\rm d}$. Here, $\delta c_{\rm d}$ represents the subset of terms present in $\delta c$ that are capable
 of biasing the dicosmic fit. 

In the following, we set $a_1 = 0$ without loss of generality since $a_1$ represents a correction to a global scale factor on curvature. 
\subsection{Analysis of $\Delta^+_c$}
\label{appendixBplus}
\hspace*{0.06in}
The $\Delta^+_c$ observable (Eq.~\ref{deltaPlusEquation}) is an expansion in small coefficients $a_n$ and $b_n$. 
 Consider a term in $\delta c$ of the form $a_nc^n$ (where $n$ is a positive integer) which upon substitution expands to 
 $$a_nc^n = a_n (c_{\rm d} - \delta c_{\rm d})^n \approx a_n(c_{\rm d}^n -nc_{\rm d}^{n-1}
 \delta c_{\rm d}) = a_nc_{\rm d}^n(1 -n \frac{\delta c_{\rm d}}{c_{\rm d}})$$ to first order in $\delta c$ when $n>0$. 

 As $\frac{\delta c_{\rm d}}{c_{\rm d}} < \frac{\delta c}{c_{\rm d}} \lesssim 10^{-4}$ (known from the measurements of $\Delta_{pe}$ and 
 the $Z$ boson mass~\cite{CDF2022}), the use of $c_{\rm d}$ as a proxy for $c$ allows us to infer the values of the $a_n$ coefficients with  a relative
 accuracy better than 0.01\% for $n > 0$. Replacing $a_n \to b_nq$ proves the same consequence for the $b_n$ coefficients for $n > 0$. Thus
 there is no loss of accuracy when $c_{\rm d}$ is substituted for $c$.

 Since this proof fails for $n=0$, we consider the $a_0$ coefficient explicitly. The use of $c_{\rm d}$ in 
 Eq.~\ref{deltaPlusEquation} yields, to first order in $\delta c$ as $c \to 0$, 
 $$\Delta^+_c \approx a_0 + b_1q (c_{\rm d} - \delta c) \approx a_0 (1 - b_1q) + b_1q c_{\rm d}$$ 
 which shows that the fractional inaccuracy on the $a_0$ determination is given by $b_1$, which is negligible. 

We conclude that one can  use
 $c_{\rm d}$ as a proxy for $c$ when analyzing $\Delta^+_c$ to measure the coefficients appearing in Eq.~\ref{deltaPlusEquation} with no
 loss of accuracy. 
 These conclusions have been verified by a Monte Carlo study of this observable. 
\subsection{Analysis of $\Delta^-_c$}
\label{appendixBminus}
\hspace*{0.06in}
The use of $c_{\rm d}$ in Eq.~\ref{deltaMinusEquation} yields, to first order in $\delta c$ as $c \to 0$,
$$  \Delta^-_c \approx b_0q + c = b_0q + (c_{\rm d} - \delta c_{\rm d}) $$
Subtracting the term linear in $c_{\rm d} $ yields, as $c \to 0$, 
$$\Delta^-_c - c_{\rm d} = b_0q [1-1 + {\cal O}(b_2c)] \propto b_0 b_2c$$
This observable is suppressed by the $b_2$ factor. Furthermore, it cancels in the analysis of $| \Delta^-_c - c_{\rm d}|$ if linear in $c$
  and is absorbed in $a_1$ if linear in $|c|$. We conclude that $b_0$ cannot be constrained by the analysis of the $\Delta^-_c$ observable. 

Next, consider a term in $\delta c$ of the form $b_rq|c|^r$ (where $0<r<1$ is a positive fraction). Without loss of generality we consider
 $c > 0$ since the calculation will be mirrored for $c<0$. Therefore 
$$  \Delta^-_c \approx b_rc^r  + c = b_rc^r + (c_{\rm d} - b_rc^r) \approx c_{\rm d} \implies \Delta^-_c - c_{\rm d}  \approx 0$$

Similarly, consider a term in $\delta c$ of the form $b_2qc^2$. For $c > 0$, 
$$  \Delta^-_c \approx b_2c^2  + c = b_2c^2 + (c_{\rm d} - b_2c^2) \approx c_{\rm d} \implies \Delta^-_c - c_{\rm d}  \approx 0$$
and the same conclusion is reached for $c<0$ by replacing $b_2 \to -b_2$. 

By extension, none of the terms in $\delta c$ that show up in the  $\Delta^-_c$ observable as a function of $c$ can be extracted 
 by studying $\Delta^-_c$ as a function of $c_{\rm d}$. 

The reason that the coefficients in Eq.~\ref{deltaMinusEquation} cannot be extracted by analyzing $\Delta^-_c$ versus $c_{\rm d}$ is 
 that $\Delta^-_c$ represents the average of the curvature measurements of the incoming (corrected for direction of propagation) and 
 outgoing legs of the cosmic ray trajectory, and $c_{\rm d}$ represents a similar (constrained) average curvature of the entire trajectory. 
 Thus $\Delta^-_c \cong c_{\rm d}$ and the analysis of $\Delta^-_c$ versus $c_{\rm d}$ is blind to the coefficients in 
 Eq.~\ref{deltaMinusEquation}. 
\section{Single-wire bias on curvature}
\label{appendixWireBias}
\hspace*{0.06in}
A curvature bias $\delta c$ is related to a sagitta ($d$) bias as $d \sim (\delta c) l^2$ where $l$ is the length of the track. 
The typical alignment accuracy of the wires is 1~$\mu$m~\cite{cosmicAlignment,CDF2022}. For  
 $d = 1~\mu$m and $l = 1$~m, $\delta c \sim 10^{-8}~ {\rm cm}^{-1} = 5~{\rm PeV}^{-1}$ for the 1.4~T magnetic field of CDF. At $p_T \sim
 40$~GeV, this curvature bias translates to a momentum bias of 200~ppm. Since a sagitta bias this large would materialize at the maximum 
 rate of 1\permil\ (Sec.~\ref{sec:COTperf}), the maximum momentum calibration bias would be 200~ppm$\times$1\permil\ or 0.2~ppm. Any
 difference of this rate between positive and negative tracks would translate to a charge-averaged mass bias; Fig.~\ref{fig:COThits} 
 suggests an additional suppression factor of 5, i.e. an upper bound on a mass bias of 40~ppb. This bound is conservative because
 a single-hit sagitta does not translate into a track sagitta given the large number of hits.

A similar calculation based on the superlayer efficiency variation of (at most) 0.1\permil\ (Sec.~\ref{sec:COTperf}) provides an even more
 stringent (and more realistic) bound of 4~ppb on a mass bias due to possible curvature-dependent pattern-recognition effects in the 
 $c \to 0^\pm$ limit.   
\section{A model for $b_0q$ term in curvature response}
\label{appendix:b0Model}
\hspace*{0.06in}
The difficulty of creating a COT  model that will generate a $b_0q$ term in the curvature response is that as $c \to 0$ there is 
 no information available to the COT on which  a charge-dependent effect can be based. This observation leads to the conclusion that,
 in this limit, the particle's charge can only be known to the particle itself. Therefore such a discontinuous response can only
 be generated by the particle's local interactions and not by the COT. 

 An example of a particle-interaction  model is an energy loss that grows quadratically with energy, $p_T^{\rm loss} = b_0 p_T^2$. Therefore,
$$\delta c = \frac{q}{p_T - b_0 p_T^2} - \frac{q}{p_T} = \frac{q}{p_T} (\frac{1}{1-b_0p_T} -1) 
 = \frac{q}{p_T} \frac{b_0p_T}{1-b_0p_T} \approx b_0q$$
where energy conservation requires the fractional energy loss to be within the range 
 $0 \leq b_0p_T < 1$. In practice, the consistency of the $m_Z$ measurement 
 using the $J/\psi$ and $\Upsilon$-based calibrations~\cite{CDF2022} implies that $b_0p_T < 100$~ppm as shown in 
 Sec.~\ref{sec:singularImpact}. Thus the $b_0p_T$ term can be neglected in the denominator and this energy-loss model induces the $b_0q$ term with $b_0 > 0$. 

 The reasoning is that, since the $p_T$ always reduces independent of charge, the change in curvature has the same sign as the charge,
 even as $c \to 0$, because the particle's charge is conserved. We also note that this model cannot generate $b_0 < 0$ because it is 
 impossible to mimic a particle gaining energy during propagation. 

 Synchrotron radiation off a charged particle in a magnetic field causes an energy loss per revolution that is proportional to
  $\gamma^4/R$, where $\gamma = E/m$ is the energy to mass ratio, i.e. the Lorentz boost factor, and $R$ is the radius of the 
 particle's circular arc~\cite{jackson}. As $R \propto E$ for a fixed magnetic field and polar angle, the reduction in the track $p_T$ is 
 $p_T^{\rm loss} \propto E^3$ per revolution. The fraction of a revolution executed inside the COT is 
 $l/(2 \pi R)$ where $l$ is the radius of the COT. Scaling $p_T^{\rm loss}$ by this fraction yields the result $p_T^{\rm loss}
 \propto E^3/R \propto E^2 \propto p_T^2$, or $p_T^{\rm loss} = b_0 p_T^2$. For a muon of $p_T = 40$~GeV, $p_T^{\rm loss}$ evaluates to
 3 meV (millielectron Volt)~\cite{jackson}, corresponding to a vanishingly small $b_0p_T = 0.0001$~ppb. 

If we imagine an energy loss model of the form $p_T^{\rm loss} = b_r p_T^{2-r} = - \delta p_T$, it leads to 
$$b_r p_T^{-r} = - \frac{\delta p_T}{p_T^2} =  q \delta c = b_r |c|^r \implies \delta c = q b_r |c|^r $$

As shown in Sec.~\ref{sec:singularImpact}, positive values of $b_0$ or $b_r$ bias the measured mass to lower values, as expected from 
 an unanticipated process of energy loss from the leptons. 

\end{document}